\documentclass[format=acmsmall, review=false, screen=true]{acmart}

\usepackage{booktabs} 
\usepackage{graphicx}
\usepackage{subfigure}
\usepackage{verbatim}
\usepackage{amsmath,epsfig}
\usepackage{setspace}
\usepackage{epstopdf}
\usepackage{framed}
\usepackage{listings}
\usepackage{placeins}
\usepackage{multirow}
\usepackage{commath}
\usepackage{array}
\usepackage[shortlabels]{enumitem}

\usepackage[ruled]{algorithm2e} 

\SetAlFnt{\small}
\SetAlCapFnt{\small}
\SetAlCapNameFnt{\small}
\SetAlCapHSkip{0pt}
\IncMargin{-\parindent}

\DeclareMathOperator*{\argmax}{\arg\!\max}

\setcopyright{acmlicensed}

\acmDOI{0000001.0000001}

\newcommand{\mxprd}{\ensuremath{\mathit{prd}}}
\newcommand{\Config}{\ensuremath{\mathit{c}}}
\newcommand{\mxcns}{\ensuremath{\mathit{cns}}}
\newcommand{\mxsrc}{\ensuremath{\mathit{src}}}
\newcommand{\mxsnk}{\ensuremath{\mathit{snk}}}

\newcommand{\mxvect}{\ensuremath{\mathit{vect}}}
\newcommand{\mxth}{\ensuremath{\mathit{Th}}}
\newcommand{\mxin}{\ensuremath{\mathit{in}}}
\newcommand{\mxout}{\ensuremath{\mathit{out}}}

\newcommand{\mxbuf}{\ensuremath{\mathit{buf}}}

\newcommand{\mxmpl}{\ensuremath{\mathit{mp}}}

\newcommand{\mxgcd}{\ensuremath{\mathit{gcd}}}
\newcommand{\mxmbr}{\ensuremath{\mathit{mbr}}}
\newcommand{\mxscore}{\ensuremath{f_s}}

\newcommand{\mxtmsv}{\ensuremath{\mathit{tmsv}}}
\newcommand{\mxtmsvpb}{\ensuremath{\mathit{tmsvpb}}}
\newcommand{\mxnextvd}{\ensuremath{\mathit{nextVCD}}}

\newcommand{\mxsyn}{\ensuremath{\mathit{syn}}}
\newcommand{\mxcfo}{\ensuremath{\mathit{cfo}}}
\newcommand{\mxrcp}{\ensuremath{\mathit{rcp}}}
\newcommand{\mxfft}{\ensuremath{\mathit{fft}}}
\newcommand{\mxdmp}{\ensuremath{\mathit{dmp}}}

\newcommand{\mxcp}{\ensuremath{\mathit{cp}}}
\newcommand{\mxsgm}{\ensuremath{\Sigma}}
\newcommand{\mxomegabuf}{\ensuremath{\Omega_\mathit{buf}}}
\newcommand{\mxallowable}{\ensuremath{\mathit{alwb}}}
\newcommand{\mxprofiled}{\ensuremath{\mathit{profiled}}}
\newcommand{\mxvisited}{\ensuremath{\mathit{visited}}}

\newcommand{\mxniav}{$N$-candidates IAV}

\begin{document}
\title[]{Memory-constrained Vectorization and Scheduling of Dataflow 
Graphs for Hybrid CPU-GPU Platforms}  

\author{Shuoxin Lin}
\affiliation{%
  \institution{University of Maryland}
  \city{College Park}
  \state{MD}
  \postcode{20742}
  \country{USA}}
\email{slin07@umd.edu}
\author{Jiahao Wu}
\affiliation{%
  \institution{University of Maryland}
  \city{College Park}
  \state{MD}
  \postcode{20742}
  \country{USA}}
\email{jiahao@umd.edu}
\author{Shuvra S. Bhattacharyya}
\affiliation{%
 \institution{University of Maryland}
  \city{College Park}
  \state{MD}
  \postcode{20742}
  \country{USA}}
\affiliation{%
 \institution{Tampere University of Technology}
  \city{Tampere}
  \country{Finland}}
\email{ssb@umd.edu}

\begin{abstract}


The increasing use of heterogeneous embedded systems with multi-core CPUs and
Graphics Processing Units (GPUs) presents important challenges in effectively
exploiting pipeline, task and data-level parallelism to meet throughput
requirements of digital signal processing (DSP) applications. Moreover, in the
presence of system-level memory constraints, hand optimization of code to
satisfy these requirements is inefficient and error-prone, and can therefore,
greatly slow down development time or result in highly underutilized processing
resources.  In this paper, we present vectorization and
scheduling methods to effectively exploit multiple forms of parallelism for
throughput optimization on hybrid CPU-GPU platforms, while conforming to
system-level memory constraints. The methods operate on synchronous dataflow 
representations, which are widely used in the design of embedded systems
for signal and information processing. We show that our novel methods can
significantly improve system throughput compared to previous vectorization and
scheduling approaches under the same memory constraints. In addition, we
present a practical case-study of applying our methods to significantly improve
the throughput of an orthogonal frequency division multiplexing (OFDM) receiver
system for wireless communications.  


\end{abstract}

%
%


%
%

\keywords{Dataflow models, design optimization, 
heterogeneous computing, signal processing systems, software synthesis}

\maketitle

\renewcommand{\shortauthors}{S. Lin et al.}

\section{Introduction}
\label{sec:introduction}

Heterogeneous multiprocessor platforms are of increasing relevance in the
design and implementation of many kinds of embedded systems. Among these
platforms, {\em heterogeneous CPU-GPU platforms (HCGPs)}, which integrate
multicore central processing units (CPUs) and graphics processing units (GPUs),
have been shown to significantly boost throughput for many applications.
System-level performance optimization requires efficient utilization of both
CPU cores and GPUs on HCGPs.  In embedded system designs, multiple system
constraints must be met including memory, latency or cost requirements. Manual
performance tuning on a case-by-case suffers from inefficiency and can lead to
highly sub-optimal solutions.  When system constraints or the target platforms
are changed, the designer often needs to repeat the same process, which further
reduces development productivity, and increases the chance of introducing
implementation errors.  Therefore, methods for HCGPs that are based on
high-level models, and systematically explore parallelization opportunities are
highly desirable. 

Dataflow models provide high-level abstractions for specifying, analyzing and 
implementing a wide range of embedded system applications 
(e.g., see~\cite{bhat2013x1}). 
A dataflow graph is a directed graph $G=(V,E)$ with a set of vertices 
({\em actors}) $V$ and a set of edges $E$. An actor $v \in V$ represents a 
computational task of arbitrary complexity.  An edge $e=(u,v) \in E$ represents 
a first-in, first-out (FIFO) buffer that stores data values 
as they are produced by $u$ and consumed by $v$.  These data values are called 
{\em tokens}, and represent the basic unit of data that is processed by actors. 
When an actor {\em fires}, it consumes tokens from its input edges, executes 
its associated task, and produces tokens on its output edges. 

\begin{figure}[]
\centering
\subfigure[]{
  \includegraphics[height=0.6in]{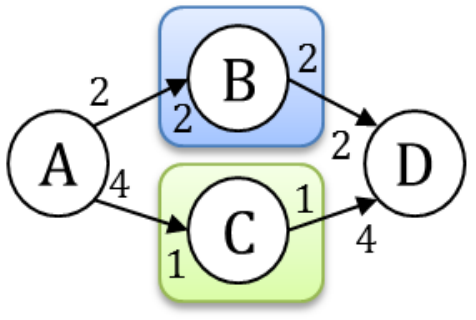}
  \label{fig:para-a}
}
\subfigure[]{
  \includegraphics[height=0.6in]{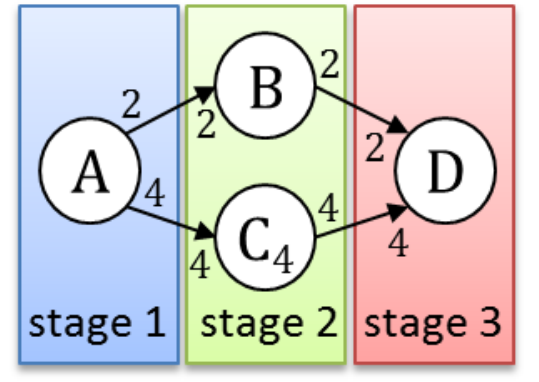}
  \label{fig:para-b}
}
\subfigure[]{
  \includegraphics[height=0.6in]{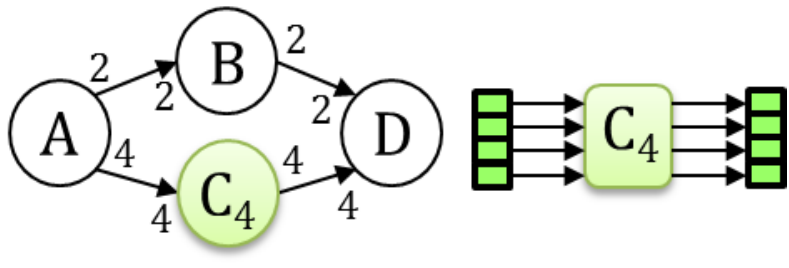}
  \label{fig:para-c}
}

\caption{An illustration of parallelism expressed using SDF graphs: 
(a) task parallelism, (b) pipeline parallelism, (c) data parallelism 
based on vectorization. }
\label{fig:para}
\end{figure}

{\em Synchronous dataflow (SDF)} is a specialized form of dataflow in which the
numbers of tokens produced and consumed on each edge are constant across all
firings of its source and sink actors~\cite{lee1987x1}. These two numbers are
called the {\em production rate} and {\em consumption rate} of an edge.
Generally, the production rate and consumption rate of an SDF edge can take on
any positive integer value. SDF graphs are powerful tools for analyzing and
optimizing important system-level metrics, including memory requirements,
latency, and throughput. Additionally, SDF graphs naturally expose pipeline,
task and data parallelism across distinct actors and distinct firings of the
same actor, as illustrated in Figure~\ref{fig:para}. Pipeline and task
parallelism can be exploited by assigning actors on different cores or
processors (Figure~\ref{fig:para-a} and \ref{fig:para-b}, while exploitation of
data parallelism can be enhanced by vectorization of actors such that different
sets of tokens are processed by the same actor concurrently on data-parallel
hardware (Figure~\ref{fig:para-c}). 

GPUs in HCGPs accelerate computational tasks by supporting large-scale data
parallelism with hundreds or thousands of SIMD (single instruction multiple
data) processors. GPUs can achieve high throughput gain over CPUs when parallel
data is abundant. However, when parallel data is insufficient, GPU performance
can be worse compared to CPU cores. For an SDF graph, a
sufficient amount of parallel data may not be present to effectively utilize a
GPU. In this case, vectorization can be of great utility in improving the
degree of exposed data parallelism, and the effective utilization of GPU
resources.  However, previous research on scheduling and software synthesis
from SDF graphs has focused largely on task and pipeline parallelism, therefore
providing inadequate support of GPU-targeted design flows.  The developments
in this paper are intended to address this gap.

In general, the average time required for an actor firing scales differently in
terms of the vectorization factor between a CPU and GPU.  Additionally,
overheads involving interprocessor communication and synchronization can limit
or even negate performance gains achieved through vectorization. Thus,
effective throughput optimization for HCGPs requires rigorous joint
consideration of vectorization and scheduling.  

In this paper, we develop integrated vectorization and scheduling (IVS)
techniques for software synthesis targeted to HCGPs. These techniques jointly
consider vectorization and scheduling for thorough optimization of SDF graphs.
We refer to this problem of joint vectorization and scheduling as the SDF {\em
vectorization-scheduling throughput optimization} ({\em VSTO}) problem, or
simply as ``VSTO''.  Our contribution is summarized as follows. First, we
formally present the VSTO problem for HCGPs.  Second, we develop a set of novel
vectorization and scheduling techniques for VSTO under memory constraints.
Third, we propose a new scheduling strategy called $\mxsgm$-scheduling that is
effective for mapping dataflow actors on heterogeneous computing platforms.
Finally, we demonstrate our approaches to VSTO by applying them to a large
collection of synthetic, randomly-generated dataflow graphs and an Orthogonal
Frequency Division Multiplexing (OFDM) receiver.

\section{Related Work}
\label{sec:related}

SDF throughput analysis under resource constraints using explicit state space
exploration has been studied in~\cite{gham2006x1}.  In~\cite{park2010x1}, the
authors present a scheduling algorithm for SDF graphs that applies static
topological analysis and vectorization to improve SDF throughput and memory
usage on shared-memory, multicore platforms.  In~\cite{chen2012x1}, a buffer
optimization technique for pipelined, multicore schedules is discussed. 

Earlier work on SDF vectorization has focused on throughput optimization for
single-processor implementation on programmable digital signal processors, and
more recently, on multicore implementation. SDF vectorization techniques to
maximize throughput for single-processor implementation were first developed
in~\cite{ritz1993x1}. In~\cite{ko2008x1}, the authors presented methods to
construct vectorized, single-processor schedules that optimize throughput under
memory constraints. In~\cite{hsu2011x1}, the authors presented techniques for
maximizing throughput when simulating SDF graphs on multicore platforms.
These techniques simultaneously optimize vectorization, inter-thread
communication, and buffer memory management. In these works, SIMD
architectures are not involved, and
vectorization is applied to reduce synchronization
overhead and context switching rather than to exploit data-parallelism. 

Various studies have targeted automated exploitation of parallelism to map
dataflow models onto heterogeneous computing platforms. Design tools that
exploit various forms of parallelism using CUDA or OpenCL have been developed
in~\cite{cicc2013x1,lund2015x1,scho2013x1}. These tools assume that
vectorization has been specified by the designer, and map an actor onto a GPU
whenever a GPU-accelerated implementation of the actor is available.  For such
actors, these tools do not take into account the possibility that CPU-targeted
execution may be more efficient. In~\cite{zaki2013x3}, SDF graphs are
automatically vectorized, transformed to single-rate SDF graphs, and then
scheduled using Mixed-Integer Programming techniques.  However, this approach
does not take memory constraints into account. Intuitively, a single-rate
SDF graph is one in which all actors are fired at the same average rate.
This concept is discussed in more detail in Section~\ref{sec:background}.

When SDF graphs are converted to single-rate graphs, they can be scheduled in
the same way that task graphs are scheduled in programming environments such as
StarPU~\cite{augo2011x1}, FastFlow~\cite{goli2012x1}, and
OmpSS~\cite{dura2011x1}. These environments support run-time task graph
scheduling and parallelization on hybrid CPU-GPU platforms.  StarPU, for
example, uses the {\em Heterogeneous Earliest Finish Time} ({\em HEFT})
heuristic to schedule tasks on HCGPs. However, these programming models cannot
directly be applied to multirate SDF graphs; a designer must manually vectorize
the graph and convert it to a single-rate SDF graph before working with it in
such environments. In addition to requiring such manual transformation, this
process limits the flexibility in vectorization and scheduling for SDF
execution, which can lead to inefficient memory usage and execution time
performance. 

Dataflow models can be used at arbitrary levels of abstraction in computing
systems, and hence compilation optimization of dataflow programs is also
investigated at various levels of abstraction. For example, the works
in~\cite{udup2009x1,hagi2011x1} focus on improving the throughput of GPU
kernels that are represented by dataflow graphs. The aim of those works is to
generate high-performance GPU kernel code through better utilization of
on-device resources. In contrast, the methods introduced in this paper focus on
optimizing the mapping of coarse-grain, system-level dataflow models onto
CPU-GPU platforms, where each actor can encompass a computational task of
arbitrary complexity, and can encapsulate one or multiple kernels.

In this work, we go beyond the previous works by jointly considering SDF
vectorization and scheduling for HCGPs under memory constraints.  To our
knowledge, our work is the first to take memory constraints into account in the
context of SDF vectorization and scheduling for heterogeneous computing
platforms.  Our methods are not restricted to single-rate SDF graphs, and are
capable of deriving efficient, memory-constrained vectorization configurations.
The techniques in this paper are developed in the
DIF-GPU Framework, which was presented in~\cite{lin2016x1}. DIF-GPU
incorporates techniques for minimizing runtime overhead through compile-time
scheduling and incorporation of carefully-designed protocols for interprocessor
communication.

\section{Background}
\label{sec:background}

The HCGPs that we target in this paper consist of one multi-core CPU and one
GPU each. This class of multicore architectures is widely used in
embedded systems. In our targeted class of HCGPs, we refer to the CPU as the
{\em host}, as it controls overall execution flow and manages the associated
GPU, and we refer to the GPU as the {\em device}. The device receives
instructions and data from the host. 

Additionally, in the target architecture, there exists a context transfer
overhead when an application's execution path switches between CPU cores and a
GPU. This overhead can include the time for interprocessor communication and
synchronization, context switching, and transferring data from one memory
address to another. Although most existing embedded HCGPs provide shared
physical memory, this context transfer overhead can still be significant, and
in general varies from one architecture / application to
another~\cite{greg2011x1}. We refer to such context transfer overhead as {\em
host-to-device (H2D)} or {\em device-to-host (D2H)} context transfer, depending
on the direction.

Given an SDF graph $G = (V, E)$ and an actor $v \in V$, we denote the sets of
input and output edges of $v$ as $\mxin(v)$ and $\mxout(v)$, respectively.
Given an edge $e \in E$, we denote the source and sink actors of $e$ by
$\mxsrc(e)$ and $\mxsnk(e)$, respectively.  We denote as $\mxprd(e)$ the number
of tokens produced onto $e$ by each firing of $\mxsrc(e)$, and similarly, we
denote as $\mxcns(e)$ the number of tokens consumed from $e$ by each firing of
$\mxsnk(e)$. 

\begin{figure}[]
\centering
\subfigure[]{
  \includegraphics[width=2in]
  {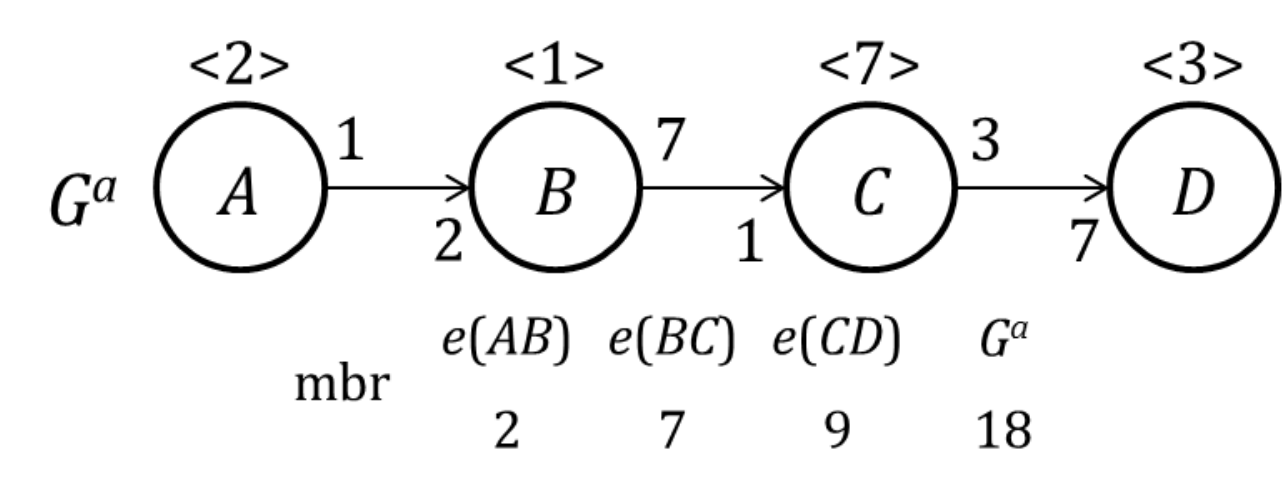}
  \label{fig:ex-a}
}
\subfigure[]{
  \includegraphics[width=2in]
  {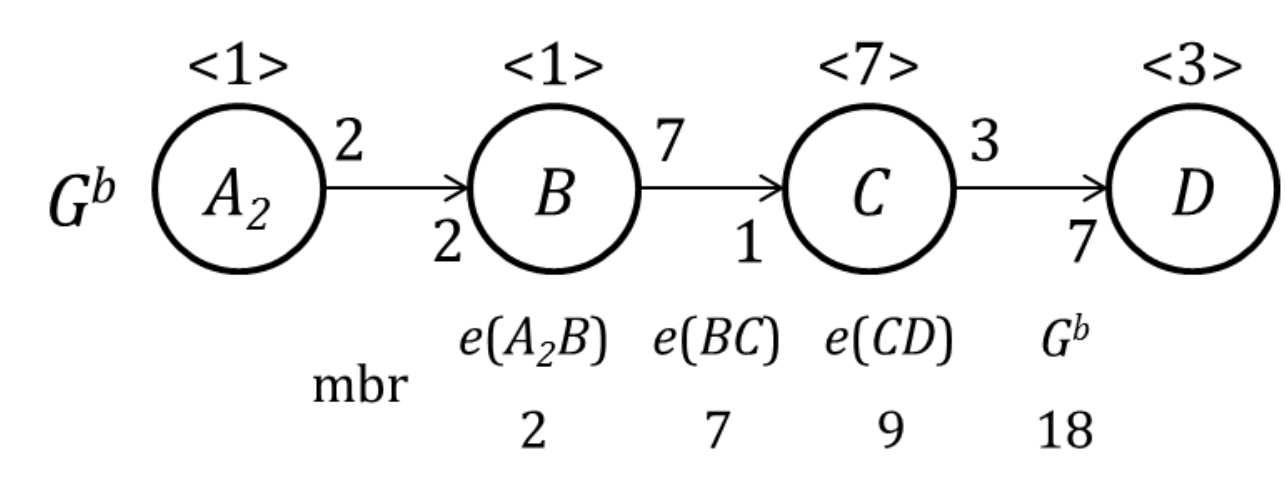}
  \label{fig:ex-b}
}
\subfigure[]{
  \includegraphics[width=2in]
  {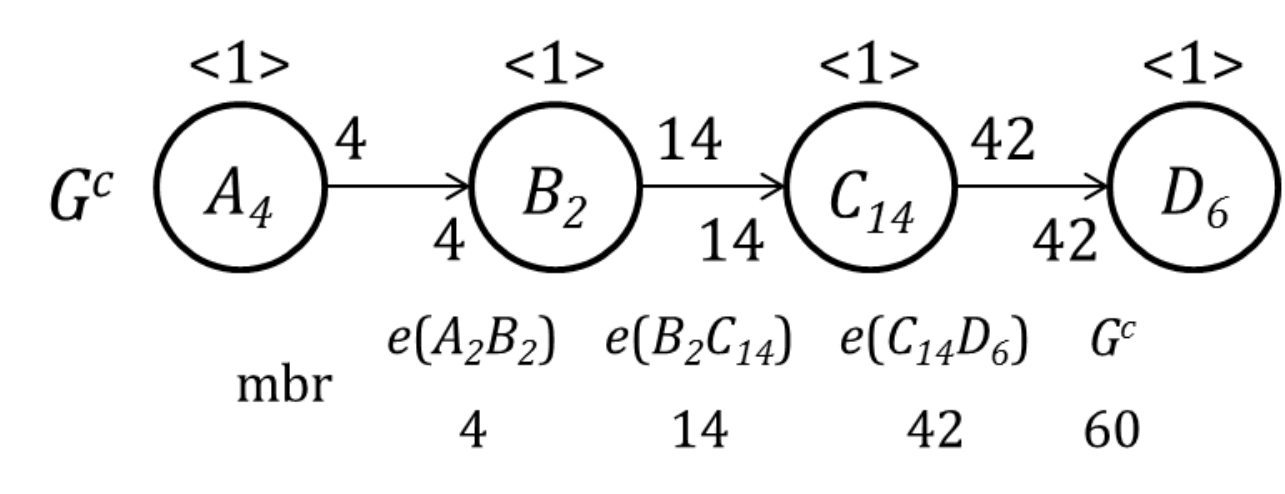}
  \label{fig:ex-c}
}
\subfigure[]{
  \includegraphics[width=2in]
  {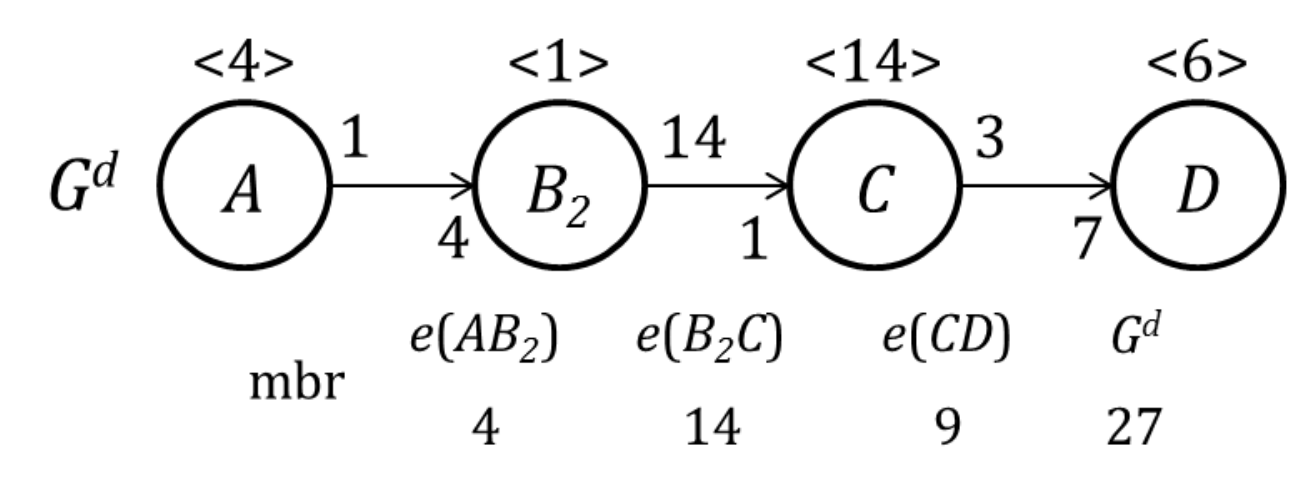}
  \label{fig:ex-d}
}

\caption{An example of vectorization and minimum buffer requirements. 
(a) Original 
graph. (b) Actor-level Vectorization of $A$ by 2. (c) Graph-level Vectorization 
with $\beta = 2$. (d) Actor-level Vectorization of $B$ by 2. }
\label{fig:ex}
\end{figure}

Signal processing systems represented as SDF graphs are often required to be
executed indefinitely --- that is, iterated through a number of iterations for
which no useful bound is known in advance. To support such indefinite
execution, the concepts of consistency and periodic schedules in SDF graphs are
important~\cite{lee1987x1}.  An SDF graph is {\em consistent} if it has a {\em
periodic schedule}, which is a sequence of actor executions that does not
deadlock, fires each actor at least once, and produces no net change in the
number of tokens on each edge.  Consistent SDF graphs can be executed
indefinitely with finite buffer memory requirements.  Furthermore, for each
actor $v \in V$ in a consistent SDF graph $G = (V, E)$, there is a unique {\em
repetition count} $q(v)$, which gives the minimum number of firings of $v$ in a
periodic schedule. We call a set of actor firings in which each actor $v$ fires
exactly $q(v)$ times an {\em iteration} of $G$.  Figure~\ref{fig:ex-a} shows an
SDF graph example, where each repetition count is denoted as $<q(v)>$ above the
corresponding actor $v$. In this example, $\mxprd(e_{AB}) = 1$, $\mxcns(e_{AB})
= 2$, $q(A) = 2$, and $q(C) = 7$. 

If $q(v) = 1$ for every actor $v \in V$, then $G$ is called a {\em single-rate}
SDF graph, as shown in Figure~\ref{fig:ex-c}. Because each actor needs to fire
only once to complete an iteration of $G$, single rate SDF graphs can be
scheduled the same way as {\em task graphs} (e.g., see~\cite{zaki2013x3}).  In
a task graph, nodes represent computational tasks, and edges represent
dependencies associated with pairs of nodes without any specific data structure
implied for inter-actor communication. A wide variety of algorithms have been
developed for scheduling task graphs onto multiprocessor systems (e.g.,
see~\cite{srir2009x1}). 

For implementation of $G$, we assume a static buffer allocation model, where we
allocate a FIFO buffer of fixed, finite size (``buffer bound'') $\mxbuf(e)$ for
each edge $e \in E$.  When an actor $v$ fires, it must satisfy (1) for each
edge $e_i \in \mxin(v)$, $e_i$ contains at least $\mxcns(e_i)$ tokens, and (2)
for each edge $e_o \in \mxout(v)$, $e_o$ contains no more than $(\mxbuf(e_o) -
\mxprd(e_o))$ tokens.  When this condition is met, the actor is said to be {\em
bounded-buffer fireable}, and SDF graph execution following this rule is called
{\em bounded-buffer execution}. 

The {\em minimum buffer requirement} for an SDF graph $G$, $\mxmbr(G)$, 
is the minimum over 
all periodic schedules of the amount of memory (in units of tokens) required 
to implement the dataflow edges in a given graph (see~\cite{stui2006x1}). 
A lower bound $\mxmbr(e)$ on the minimum buffer requirement for a delayless SDF
edge $e$ can be determined by 

\begin{equation} 
\label{eq:mbr}
\mxmbr(e) = \mxprd(e) + \mxcns(e) - \mxgcd(\mxprd(e), \mxcns(e)),
\end{equation}

\noindent where $\mxgcd$ represents the greatest common divisor
operator~\cite{bhat1996x5}.  The lower bound of $\mxmbr(G)$ is the sum of
$\mxmbr(e)$ over all edges: $\mxmbr(G) = \sum_{v \in V} \mxmbr(e)$. Although
this lower bound is not always achievable, it is achievable for the dataflow
graphs in Figure~\ref{fig:ex}.

We represent the individual processors in the target multiprocessor platform as
$P = \{ p_1, p_2, \ldots, p_N\}$, where $p_1, p_2, \ldots, p_{N-1}$ represent
the available CPU cores, and $p_N$ represents the GPU.  When scheduling $G$
onto the platform, actor firings are assigned to processors to be executed. In
this context, we say that an actor $v \in V$ is {\em mapped} onto processor $p
\in P$ if all firings of $v$ are assigned to execute on $p$. 

As mentioned in Section~\ref{sec:related}, we assume in this paper that the
input SDF graphs for vectorization and software synthesis are acyclic.  Cycles
in synchronous dataflow models may impose complex constraints on what
vectorization degrees are valid for actors~\cite{ritz1993x1}. Furthermore,
cycles introduce complex trade-offs between code size and buffer memory
minimization in SDF graphs, which are also relevant to memory-constrained
vectorization problems (e.g. see~\cite{bhat1996x5}).  Third, acyclic SDF graphs
encompass a broad class of important signal processing applications, so
techniques for this class have significant practical
relevance~\cite{bhat1996x5}. Currently in our framework, we assume that actor
vectorizations are constrained only by memory, and not by cycles in the input
graph. Investigating vectorization with topological constraints caused by
cycles is an interesting direction for future work. 

\section{Problem Formulation}
\label{sec:defs}

In this section, we formally define the VSTO problem for HCGPs.  We begin by
defining the concept of actor-level vectorization.  Given a consistent SDF
graph $G=(V,E)$, and an actor $v \in V$, the {\em vectorization} of $v$ by a
{\em vectorization degree} ({\em VCD}) $b$ is defined as a transformation of
$G$ that involves the following set of operations: (1) replacing $v$ by $v_b$,
where firing $v_b$ is equivalent to $b$ consecutive firings of $v$; (2)
replacing each edge $e_i \in \mxin(v)$ by an edge $e_i'$ such that
$\mxcns(e_i') = b \times \mxcns(e_i)$ and $\mxprd(e_i') = \mxprd(e_i)$; and (3)
replacing each edge $e_o \in \mxout(v)$ by an edge $e_o'$ such that
$\mxprd(e_o') = b \times \mxprd(e_o)$ and $\mxcns(e_i') = \mxcns(e_i)$.  We
refer to the actor $v_b$ as the {\em $b$-vectorized actor} of $v$, and the {\em
transformed graph} that results from the vectorization operation as $\mxvect(G,
v, b)$.  For example, in Figure~\ref{fig:ex}, $G^b = \mxvect(G^a, A, 2)$.  The
definition of vectorization that we adopt here corresponds to a dataflow graph
transformation that is consistent with the vectorization
concept introduced by Ritz et
al.~\cite{ritz1993x1}, as opposed to the aggregation of basic operations that
corresponds to vectorization in compilers for procedural programming languages.

If $G$ is a consistent, acyclic SDF graph, then 
$\mxvect(G, v, b)$ is also consistent for any $v \in V$, and any 
positive integer $b$.
However, in this work, we restrict the set of allowable 
vectorization degrees to the set $\mxallowable(G, v)$, which
is defined as 

\begin{equation}\label{eq:allowable} 
\mxallowable(G, v) = \{n \in \{1, 2, \ldots\} \mid 
(n \mathrm{\ is\ a\ factor\ of\ } q(v)) \mathrm{\ or \ }
(n \mathrm{\ is\ a\ multiple\ of\ } q(v))\}.
\end{equation}

\noindent Equation~\ref{eq:allowable} refers specifically to positive integer 
factors and multiples. For example, if $q(v) = 8$, then
$\mxallowable(G, v) = \{1, 2, 4, 8, 16, 24, \ldots\}$. 
Vectorization of an actor $v$ that is restricted to $\mxallowable(G, v)$ 
enables fast derivation of repetition counts for
$G' = \mxvect(G, v, b)$, which in turn facilitates incremental vectorization techniques,
where actors are selected for vectorization one at a time according to specific
greedy criteria.  In particular, if $b$ is a factor of $q(v)$, then $q(G', v) =
q(G, v) / b$ , while the repetition counts of all other actors are unchanged.
Similarly, if $b$ is a multiple of $q(v)$, then $q(G', v) = 1$, while for any
other actor $u \neq v$, $q(G' u) = bq(G, u)/q(G, v)$.  In
Section~\ref{sec:vect-sched}, we discuss specific techniques for incremental
vectorization that apply these forms of repetition count updates.

On HCGPs, vectorized actors can exploit SIMD processors such as GPUs to execute
multiple firings of the same actor in parallel. Note that although parallel
processing of tokens cannot in general be applied easily to stateful actors,
vectorization may still benefit dataflow execution by reducing overheads
associated with inter-processor communication, synchronization and context
switching. In the presence of memory constraints, there are limits to the
amount of vectorization that can be applied. For example, as we can see in
Figure~\ref{fig:ex}, vectorizing $A$ (Fig.~\ref{fig:ex-a}) and vectorizing $B$
(Fig.~\ref{fig:ex-d}) by 2 results in different increases to the minimum buffer
requirement.

To represent SDF graphs with vectorized actors and their relationships with the
original graphs, we define {\em vectorized SDF graphs} ({\em VSDFs}) as
follows. 

\begin{definition}
\label{def:vect}
Suppose that $G = (V,E)$ is a consistent SDF graph, $b_v \in \mxallowable(v,
G)$ is a VCD for each $v$, and $B = \{(v,b_v) \mid v \in V\}$.  Then the {\em
$B$-vectorized SDF graph of $G$} is defined as $G_B = (V_B, E_B)$, where (1)
each $v_{B} \in V_B$ is the $b_v$-vectorized actor of $v$, (2) each edge $e_B =
(x_B, y_B)$ in $G_B$ is derived from the corresponding edge $(x, y) \in E$, and
(3) for each $e_B=(u_B, v_B) \in E_B$, $\mxprd(e_B) = b_u \times \mxprd(e)$,
and $\mxcns(e_B) = b_v \times \mxcns(e)$, where $e=(u,v)$.  
\end{definition}

The vectorized graph $G_B$ is an SDF graph. We define a restricted form of
vectorization, called {\em graph-level vectorization (GLV)}, in which a common
``repetitions vector multiplier'' $\beta \in \{1, 2, \ldots\}$ is used for all
actors in the input graph. That is, $b_v = \beta \times q(G, v)$ for all $v \in
V$.  In this context, we refer to $\beta$ as the {\em graph vectorization
degree} ({\em GVD}).  Under GLV , $G_B$ is a single-rate SDF graph.  However,
vectorization does not need to be confined to GLV.  We refer to this more
general form of vectorization, as {\em actor-level vectorization} ({\em ALV}).
For example, Figure~\ref{fig:ex-c} shows the vectorized graph that corresponds
to Figure~\ref{fig:ex-a} with GLV and $\beta=2$. Figure~\ref{fig:ex-d} shows
the vectorized graph that results from applying ALV to Figure~\ref{fig:ex-a}
with $b_B=2$. 


As discussed in Section~\ref{sec:related}, the conventional approach to solving
VSTO involves 3 steps: (1) the designer or design tool sets the GVD based on
memory constraints, (2) converts the SDF graph into a single-rate SDF graph
using GLV, and (3) generates a schedule using task graph scheduling methods.
Compared to ALV, GLV can require significantly larger buffers (see
Figure~\ref{fig:ex-c}).  The vectorization methods that we present in this
paper go beyond these conventional approaches by considering general ALV
solutions instead of being restricted only to GLV solutions.

For multiprocessor scheduling of ALV solutions, we introduce in this work a
general scheduling strategy, which is suitable for HCGPs, and can loosely
be viewed as
a variant of the list scheduling strategy. This variant is adapted for memory-constrained,
multiprocessor mapping of transformed graphs that result from ALV.
This 
strategy is a static scheduling strategy that
operates using compile-time estimates of actor execution times.
The general strategy is defined as follows.

\begin{definition}
\label{def:omega}
Given a consistent SDF graph $G=(V,E)$, and a multiprocessor target architecture with a set of 
processors $P$, the {\em \mxsgm-scheduling strategy} (1) statically assigns each actor $v \in V$ 
to a processor $p \in P$, (2) statically determines a buffer bound $\mxbuf(e)$ 
for each edge $e \in E$, and (3)
iteratively selects a bounded-buffer firable actor 
to fire on its assigned processor $p$ as soon as $p$ has completed
all executions.
An algorithm that
conforms to this scheduling strategy completes when all actors in $G$
have been scheduled using the iterative process of Step (3).
\end{definition}

The \mxsgm-scheduling strategy is closely related to 
the $\Omega$-scheduling strategy, which was
introduced in~\cite{hsu2011x1}. Both the
$\mxsgm$ and $\Omega$ strategies satisfy Parts
(1) and (2) of
Definition~\ref{def:omega}; the main difference is that with respect to
Part (3), $\mxsgm$-scheduling maps actors onto a finite number of processors, while $\Omega$-scheduling assumes an unlimited number of processors. 
Additionally, in our application of \mxsgm-scheduling,
we perform ALV to construct the input graph to the strategy.
In contrast, $\Omega$-scheduling in~\cite{hsu2011x1}
is applied to the original (unvectorized) SDF graph.

To determine the buffer bounds $\{\mxbuf(e)\}$ in \mxsgm-scheduling,
we apply the $\Omega$-buffering technique defined
in~\cite{hsu2011x1}. This technique derives the buffer bounds by
applying $\Omega$-scheduling, and determining the buffer bounds
to be equal to the corresponding buffer sizes $\{\mxbuf(e)\}$ that result
from $\Omega$-scheduling. We refer to the buffer bound
$\mxbuf(e)$ for each edge $e$ that is computed in this way
as the {\em $\Omega$ buffer bound for $e$}.
It is shown that $\Omega$-buffering
sustains maximum throughput for SDF graphs under $\Omega$ scheduling~\cite{hsu2011x1} so that imposing these bounds imposes no theoretical limitation
on throughput. Given an SDF graph $G = (V, E)$, we denote by 
$\mxomegabuf(G)$ the total
buffer memory cost for $G$ as determined by $\Omega$-scheduling:
$\mxomegabuf(G) = sum_{e \in E}(\mxbuf(e))$.

\begin{definition}
\label{def:relative}
Suppose that $G = (V,E)$ is a consistent SDF graph, $b_v \in \mxallowable(v,
G)$ is a VCD for each $v$, $B = \{(v,b_v) \mid v \in V\}$, $S_B$ 
is a periodic schedule for the $B$-vectorized graph $G_B$, and $T(S_B)$ is an estimate
of the time required to execute a single iteration of $S_B$. Then from
the fundamental properties of periodic SDF schedules~\cite{lee1987x1},  
we can derive a unique positive integer
$J(S_B, G)$, which we call the {\em blocking factor} of $S_B$ relative to $G$,
such that $S_B$ executes each $v \in V$ exactly $(J(S_B, G) \times q(G, v))$
times. In this context, we define the {\em relative throughput of $S_B$}
or the {\em throughput of $S_B$ relative to $G$} by the
quotient $J(S_B, G) / T(S_B)$. This metric gives the average number
of iterations of the original (unvectorized) SDF graph that is
executed per unit time by the schedule $S_B$.
\end{definition}

As an example, in Figure~\ref{fig:ex}, executing one iteration of $G^b$, $G^c$
or $G^d$ is equivalent to executing two iterations of $G^a$. Thus,
$J(S_B,G^b)=J(S_B,G^c)=J(S_B,G^d)=2$.

Intuitively, vectorization improves relative throughput when $T(S_B) < J(S_B)
\times T(S)$, where $S$ is the best available minimal-periodic (unvectorized)
schedule for $S$.  Such efficiency in the vectorized execution time $T(S_B)$
can be achieved due to improved utilization of processing resources under
carefully-optimized GLV and ALV configurations. 

A limitation of the vectorization techniques developed in this paper is that
they may increase latency, and thus, they may not be suitable for
implementations in which latency is a critical performance metric.  However, it
is envisioned that the methods developed in this paper provide a useful
foundation that can be built upon for latency-aware vectorization.
Investigating adaptations of these methods to take latency constraints into
account is an interesting direction for future work.

Based on the definitions introduced in this section, we formulate the
VSTO problem as follows.

\begin{definition} 
\label{def:vsto}
Let $G=(V,E)$ be a consistent SDF graph, and $P=\{p_1, p_2, \ldots, p_N\}$ be the
set of processors in an HCGP, where $p_1, p_2, \ldots, p_{N-1}$ represent the CPU cores,
and $p_N$ represents the GPU.  Given a total memory budget $M$ (a positive
integer), the {\em
vectorization-scheduling throughput optimization problem}, or {\em VSTO
problem} associated with $G$ and $P$ is the problem of finding a set $B$ of
vectorization degrees, and a schedule $S_B$ for $G_B = (V_B, E_B)$ such that
the throughput of $S_B$ relative to $G$ is maximized subject to $\mxomegabuf(G_B) \leq M$.
\end{definition}

We refer to a set of ordered pairs $C = \{(v,c_v) \mid (v \in V) \mathrm{\ and\
} (c_v \in \mxallowable(G, v))\}$ as an {\em ALV configuration} for $G$.  Note
that if an actor is not represented within a given ALV configuration (i.e., it
does not appear as the first element of any ordered pair in the set), then the
actor is assumed to be unvectorized (equivalent to a vectorization degree of
$1$).  Thus, the VSTO problem can be thought of as the problem of
jointly determining an ALV configuration $B$ together with a schedule for $G_B$
such that the resulting schedule optimizes throughput subject to a given buffer
memory constraint $M$.

The vectorization formulation and techniques developed in this paper assume
that each SDF edge (FIFO buffer) is implemented in a separate block of memory.
Various techniques have been developed in recent years to share memory
efficiently among edges in multirate SDF graphs 
(e.g., see~\cite{trip2013x1,desn2015x1}). Extending the techniques in this 
paper to incorporate such memory
sharing techniques is a useful direction for future work.

%




\section{Vectorization and Scheduling with Memory Constraints}
\label{sec:vect-sched}

In this section, we develop three main heuristics, called Incremental Actor
Vectorization (IAV), $N$-candidates IAV, and Mapping-Based Devectorization, for
the VSTO problem.  These three heuristics can be viewed as ``peers'' in the
sense that any one of them may be the preferable choice for a given
application.  Thus, the designer or a design tool can apply all three of these
complementary methods and select the best result for a given application. This
is how we have integrated the three heuristics in our DIF-GPU software
framework. More details on the integration with software synthesis and
associated experimental results are discussed in Section~\ref{sec:experiments}
and Section~\ref{sec:ofdm}.

\subsection{Incremental Actor Vectorization}
\label{sec:iav}

In this section, we define a general approach for searching the space of ALV
configurations that is based on selecting and vectorizing actors one at a time
using some specific greedy criteria.  We refer to this general approach as
{\em Incremental Actor Vectorization} ({\em IAV}).  Each iteration of IAV,
called an {\em IAV iteration}, involves the selection and vectorization of a
single actor. This results in a sequence of
intermediate vectorized graphs, $I_1, I_2, \ldots, I_N$,
where $I_i$ is the transformed graph that results from IAV iteration $i$,
and $N$ is the total number of iterations before IAV terminates. The approach 
is incremental in both the dimensions of actors and
vectorization degrees --- that is, each IAV iteration selects a single actor
$v$, and increases its vectorization degree to the next highest element of
$\mxallowable(G, v)$.  Given an actor $v$ that has an associated vectorization
degree $b_v$, we refer to this process of replacing $b_v$ with the next highest
element $\mathit{min}(x \in \mxallowable(v) \mid x > b_v)$ as {\em stepping up}
the vectorization of $v$ or just ``stepping up $v$''. 

In IAV, we define a ``score'' function to guide the vectorization process. At
each algorithm iteration, IAV selects an actor that has the highest score among
all actors whose stepping up would not result in a violation of the given
memory budget $M$.  Analogous to how different priority functions can be used
to select tasks in multiprocessor list scheduling (e.g.,
see~\cite{srir2009x1}), different score functions can be used to apply
different ALV criteria in IAV. This contributes to a novel design space for
development of integrated vectorization and scheduling techniques.

The specific score functions that we experiment with in this work first apply
$\mxsgm$-scheduling to generate a schedule $\mu(i)$ of the current $I_i$
(intermediate vectorized graph) onto the target HCGP $P$, and then use a
specific metric to estimate the potential ``gain'' of each candidate stepping
up operation relative to the processor assignment associated with $\mu(i)$.
Given a schedule $S$ returned by $\mxsgm$-scheduling, we define the associated
{\em processor assignment} associated with $S$ and dataflow graph $G = (V, E)$
as the function $\mxmpl_S : V \rightarrow P$ such that for each $v \in V$,
$\mxmpl_S(v)$ gives the processor to which actor $v$ is mapped according to $S$.
The initial schedule $\mu(0)$ is derived by applying $\mxsgm$-scheduling
to the input (unvectorized) graph for IAV.

Algorithm~\ref{algo:vect} shows a
pseudocode description of the IAV approach that employs this mapping-based
method of score function formulation.
In the remainder of this paper, we refer to the mapping-based
form of IAV shown in Algorithm~\ref{algo:vect} as ``\mxsgm-IAV''.

\begin{algorithm}
\SetKwData{ConfigList}{$\mathit{configs}$} 
\SetKwData{Config}{$\mathit{c}$}
\SetKwData{Mapping}{$\mathit{mp}$}
\SetKwFunction{Throughput}{throughput}
\SetKwFunction{GenMapping}{generateMapping}
\SetKwFunction{MemReq}{memSize}
\SetKwFunction{IncVec}{incrementalVectorize}
\SetKwFunction{NextVCD}{nextVCD}
\SetKwFunction{Vectorize}{vectorize}
\SetKwFunction{Score}{score}
\SetKwProg{Fn}{Function}{}{}
\Fn{$\IncVec(G,P,M)$}{
initialize $\ConfigList = \emptyset$, $G_B = G$, 
$B=\{(v,1)| v \in V\}$ \; 
\While{$\MemReq(G_B) \leq M$}{
  $\Mapping = \GenMapping(G_B, P)$\; 
  $v^* =\argmax_{v \in V} \Score(B, \Mapping, v)$\; 
  $B(v^*) = \NextVCD(v^*,b_{v^*})$ \; 
  $G_B = \Vectorize(G, B)$ \; 
  \If{$\MemReq(G_B) \leq M$}{
    $\ConfigList = \ConfigList \cup \{(B, \Mapping)\}$ \;
  }
}
\Return $\argmax_{\Config \in \ConfigList}\Throughput(G, \Config)$
}
\caption{Integrated vectorization and mapping using 
\mxsgm-IAV.}
\label{algo:vect}
\end{algorithm}

In Algorithm~\ref{algo:vect}, \verb#generateMapping# is
a placeholder for any $\mxsgm$-scheduling technique
that is applied to map
a given intermediate vectorized graph onto the targeted heterogeneous 
platform $P$. In our implementation of $\mxsgm$-IAV, we employ
a specific $\mxsgm$-scheduling technique called {\em Incremental Actor Re-assignment (IAR)} 
as the \verb#generateMapping# function. The IAR technique is discussed further
in Section~\ref{algo:map}. The function $\mxnextvd(v,b_v)$ gives 
smallest element of
$\mxallowable(v)$ that exceeds $b_v$.

The function \verb#throughput# referenced in Algorithm~\ref{algo:vect}
represents a placeholder for any function that is used to estimate the
throughput of a mapping that is generated by \verb#generateMapping# for an
intermediate vectorized graph. In our implementation of $\mxsgm$-IAV, we employ
an efficient simulation-based approach for this kind of throughput estimation.
This simulation approach is discussed further in Section~\ref{sec:throughput}.
In general, heuristic-based mapping techniques, including our techniques, do
not guarantee an optimal scheduling. It is therefore possible for the
throughput to get worse during incremental vectorization.  For this reason, we
assess the throughput of each computed configuration and then select a
configuration that results in the best throughput.

We formulate and experiment with two specific score functions in this work. We
refer to these score functions as {\em time-saving} ({\em TMSV}) and {\em
time-saving-per-byte} ({\em TMSVPB}).  The TMSV score for actor $v$
during IAV iteration $i$ is defined as largest {\em adjusted execution time
reduction} achievable (across all processors in $P$) when stepping up $v$. This
``adjusted'' time reduction is computed relative to the execution of $v$ on
$\mxmpl_{\mu(i)}(v)$, and is normalized by the vectorization degree. The units
of this adjusted time reduction are thus ``seconds per unit of vectorization''.
This score can be expressed as:

\begin{equation}
\label{eq:tmsv}
\mxtmsv(v, i) = \max_{p \in P}(
\frac{t(v,b,\mxmpl_{\mu(i)}(v))}{b} - \frac{t(v,b',p)}{b'}),
\end{equation}

\noindent where 
$b$ is the current VCD of $v$ (in IAV iteration $i$), and $b' \in 
\mxallowable(v)$ is the VCD that would result
from stepping up $v$. 
For a given actor $v$, vectorization degree $b \in \mxprofiled(v)$,
and processor $p \in P$, $t(v, b, p)$ gives the profiling-derived
estimate for the execution time 
of $v$ on $p$ with vectorization degree $b$. 

Here, we use ``profiling'' as a general term that encompasses any method for
deriving a compile-time estimate for the execution time of a vectorized actor
execution.  The specific approach to profiling that we use in our experiments
is discussed in Section~\ref{sec:experiments}.

\begin{figure}[]
\centering
\includegraphics[width=4.2in]{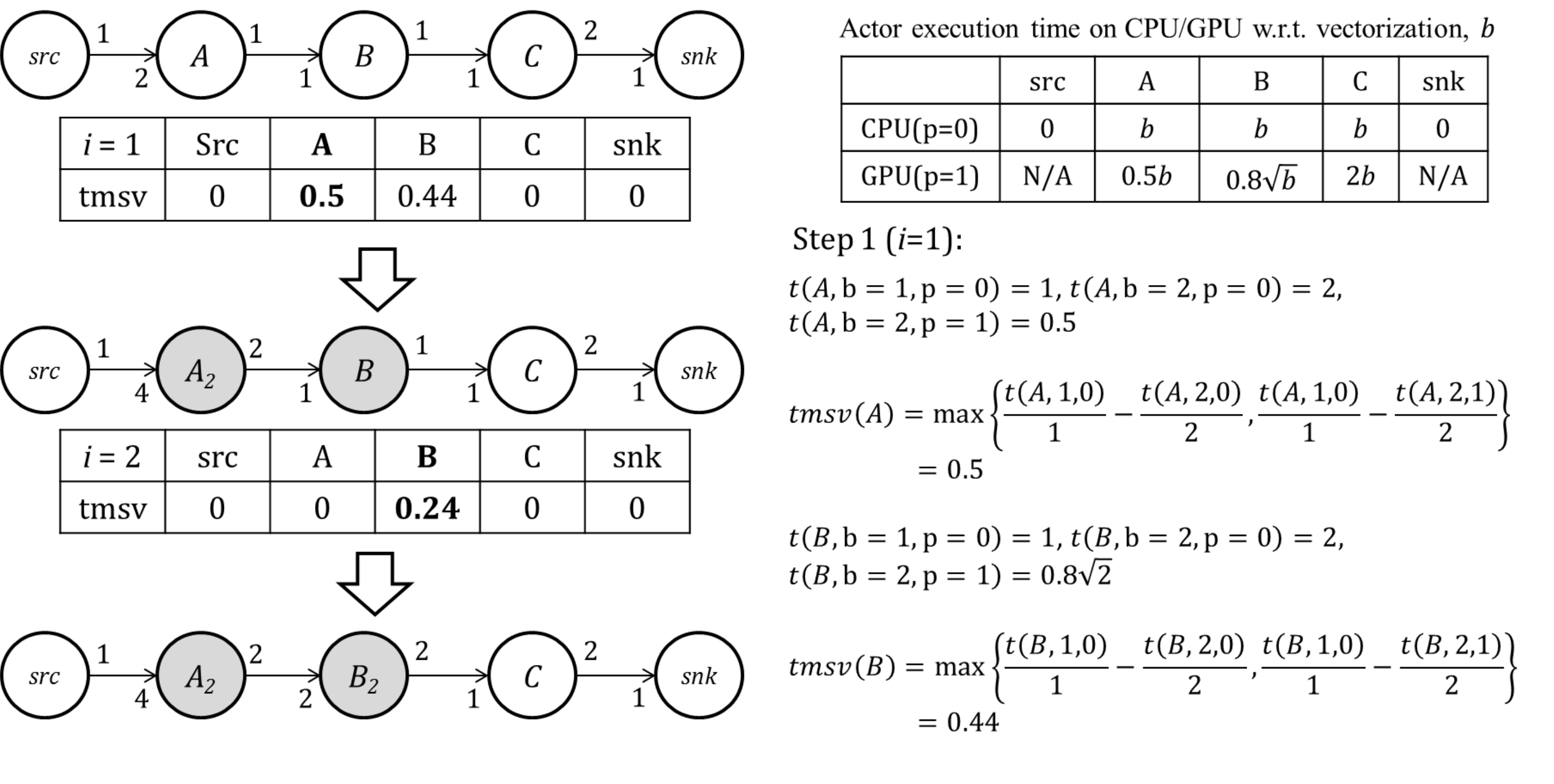}
\caption{A simple example to illustrate \mxsgm-IAV using 
the TMSV score function.}
\label{fig:vect-tmsv}
\end{figure}

Figure~\ref{fig:vect-tmsv} shows a simple example of vectorization using 
the TMSV
score function. The table in this figure provides analytical models,
in terms of the vectorization degree $v$,
that are used to derive the profiling function $t$. 
For example, the models estimate that actor $A$ requires 
approximately $(0.5 \times v)$ units of time to execute.

The IAV process begins
with an unvectorized graph and an initial mapping where all actors are mapped to
the CPU core. In the first IAV iteration ($i = 0$)
shown in Figure~\ref{fig:vect-tmsv}, $A$ has the largest TMSV score, so it is
selected, and a new mapping is generated based on the VCDs.  In the second
iteration, $B$ has the largest TMSV score, so $B$ is vectorized
(stepped up), and the mapping is
updated again. This process continues until no more vectorization
operations can be carried out without exceeding the memory budget $M$.
 
Under memory constraints, we expect that it will be more useful to consider 
the increase in buffer requirements when selecting actors for ALV. This
motivates our formulation of the TMSVPB score function. Here,
``PB'' stands for ``per byte.''
This memory-aware score function can be formulated as:

\begin{equation}
\label{eq:tmsvpb}
\mxtmsvpb(v, i) = \max_{p \in P}(
\frac{t(v,b,\mxmpl_{\mu(i)}(v))/b - t(v,b',p)/b'}{\mxomegabuf(G_{B'}) - \mxomegabuf(G_{B(i)}) + \epsilon}),
\end{equation}

\noindent where $B(i)$ represents the current ALV configuration in ILV
iteration $i$,  and $B' = B(i) - \{(v,b)\} + \{(v,b')\}$ represents the
candidate configuration that results from stepping up $v$. $\epsilon$ is 
a small constant to avoid division by 0 when 
$\mxomegabuf(G_{B'}) = \mxomegabuf(G_{B(i)})$ . Thus, the TMSVPB
function favors actors whose vectorization results in throughput improvement
without excessive increase in buffer requirements. 

\subsection{$N$-Candidates IAV}
\label{sec:niav}

Our proposed $\mxsgm$-IAV approach has two drawbacks --- (1) it selects only
one actor at each step, and (2) with the TMSV and TMSVPB score functions, the
selections are based primarily on actor execution times, and 
do not take into account
the SDF graph topology.  We alleviate the first drawback by storing
multiple vectorized-graph candidates to consider in each IAV iteration
following the very first iteration. In particular, we store 
$N$ candidate graphs that provide the highest throughput when processed
by $\mxsgm$-scheduling. Here, $N$ is a parameter that can be controlled
by the designer or tool developer.

The second drawback can be addressed by
applying $\mxsgm$-scheduling to optimize throughput over each actor for every
candidate graph. That is, for each candidate graph $Y$ that is stored, and each
actor $v$, we apply $\mxsgm$-scheduling to the transformed graph that results
from stepping up $v$ in $Y$. We then take the best result from all of these
$\mxsgm$-scheduling-based evaluations to determine the vectorization operation
that is to be applied in the associated IAV iteration. This approach results in
some increase in complexity, but has the potential to perform significantly
more thorough optimization at a relatively high level of design abstraction.

We refer to this modified \mxsgm-IAV approach as {\em $N$-candidates IAV}.  
Algorithm~\ref{algo:ncan} provides a pseudocode description of
$N$-Candidates IAV. Here, the notation $c.1$  denotes the first element of 
the ordered pair $c$, and  $\mathit{configs}[1:N]$ denotes the list that 
consists of the first $N$ elements of the list $\mathit{configs}$.
The function $\mxvisited(B')$ tests whether the vectorization 
configuration $B'$ has been examined before during operation
of the algorithm.

\begin{algorithm}
\SetKwData{ConfigList}{$\mathit{configs}$} 
\SetKwData{Config}{$\mathit{c}$}
\SetKwData{Mapping}{$\mathit{mp}$}
\SetKwData{Flag}{$\mathit{flag}$}
\SetKwData{True}{$\mathit{true}$}
\SetKwData{False}{$\mathit{false}$}

\SetKwFunction{Throughput}{throughput}
\SetKwFunction{GenMapping}{generateMapping}
\SetKwFunction{MemReq}{memSize}
\SetKwFunction{Visited}{visited}
\SetKwFunction{NCanVec}{nCandidatesVectorize}
\SetKwFunction{NextVCD}{nextVCD}
\SetKwFunction{Vectorize}{vectorize}
\SetKwFunction{Sort}{sortByThroughput}
\SetKwProg{Fn}{Function}{}{}

\Fn{$\NCanVec(G=(V,E),P,M, N)$}{
initialize $B=\{(v,1)| v \in V\}$,  
$\Mapping = \GenMapping(G, P)$,  
$\ConfigList = \{(B, \Mapping)\}$,  $\Flag = \True$ 
\While{$\Flag = \True$}{
  $\Flag = \False$ \; 
  \ForEach{$\Config \in \ConfigList$}{
    \ForEach{$v \in V$}{
     $B' = \Config.1 - \{(v, b_v)\} \cup \{(v, \NextVCD(v,b_v)\}$\;
      \If{$(\Visited(B') = \False) \mathit{and} (\mxomegabuf(G_{B'}) \leq M)$}{
        $\Mapping = \GenMapping(G_{B'}, P)$\; 
        $\ConfigList = \ConfigList \cup \{(B', \Mapping)\}$ \; 
        $\Flag = \True $, $\Visited(B') = \True$ \; 
      }
    }
  }
  \Sort(\ConfigList)\; 
  $\ConfigList = \ConfigList[1:N]$
}
\Return $\argmax_{\Config \in \ConfigList}\Throughput(G,\Config)$
}
\caption{A pseudocode description of $N$-candidates IAV.}
\label{algo:ncan}
\end{algorithm}

As with our implementation of $\mxsgm$-IAV, we employ in our implementation of
$N$-candidates IAV the IAR technique (Section~\ref{algo:map}) as the
\verb#generateMapping# function.  Similarly, our implementation of
$N$-candidates IAV incorporates the simulation-based throughput estimation
technique that is discussed in Section~\ref{sec:throughput}. This estimation
technique corresponds to the function called \verb#throughput# in
Algorithm~\ref{algo:ncan}.

Intuitively, $N$-candidates IAV is a greedy method that tries to avoid
unsatisfactory search paths by retaining multiple intermediate vectorized
graphs during each IAV iteration. Larger values for the parameter $N$ allow
more extensive design space exploration at the cost of greater running time.
When $N = 1$, $N$-candidates IAV reduces to IAV with the score function being
the estimated throughput (``throughput'') of the transformed graph that results from the
selected vectorization operation. In our implementation of $N$-candidates IAV,
we estimate throughput using simulation. This simulation approach is discussed
further in Section~\ref{sec:throughput}. In Algorithm~\ref{algo:ncan},
$\Throughput(G,\Config)$ represents the estimate of throughput that is derived
in this way for a given intermediate vectorized graph $G$ that is based on ALV
configuration $\Config$.

Other score functions can be used in $N$-candidates IAV other than throughput.
However, in our experiments, we found that among TMSV, TMSVPB, and throughput,
the throughput score function produces the best results. Investigation of other
score functions in this context is an interesting direction for future work. In
our experiments, we use $N = |V|$ as the number of candidates to be stored.  We
select $N=|V|$ so that NIAV keeps a number of candidates that scales with the
number of actors in the dataflow graph while keeping analysis time manageable. 

IAR, IAV and NIAV are all greedy-algorithm motivated heuristics based on an
evaluation metric (score function) to select vectorization choices at each
step. Investigation of other types of heuristics to further improve 
vectorization is a useful direction for future work. 

\subsection{Mapping-Based Devectorization}

$N$-candidates IAV is an incremental vectorization method that starts with
an unvectorized graph, and gradually increases the VCDs of selected actors.  In
some cases, it may be advantageous to also consider decreasing VCDs during
the optimization process. Such decreasing of VCDs can be useful to reduce
memory consumption associated with selected actors so that memory can be
dedicated to groups of other actors that provide greater throughput benefit
through vectorization. A specific form of decrease that we consider in
this section is {\em devectorization}, where an actor with 
VCD $b > 1$ is transformed to have no vectorization (VCD of unity).

\begin{figure}[]
\centering
\includegraphics[width=4in]{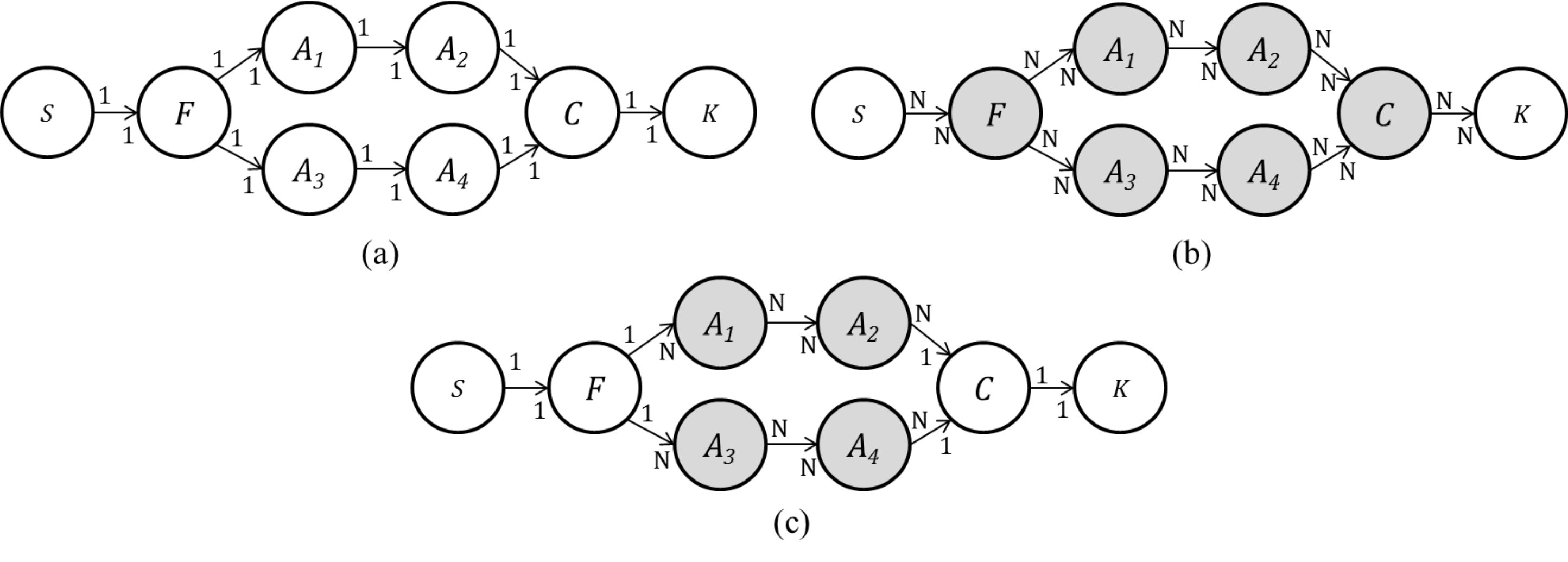}
\caption{An example that illustrates the utility of
devectorization. (a) The original graph. (b) The graph with 
$\mathrm{GVD} = N$ applied. (c) 
The graph with devectorization applied
to all CPU-mapped actors --- $C, F, K, S$.}
\label{fig:devect}
\end{figure}

Figure~\ref{fig:devect}(a) shows an example of this kind of scenario.  Here, $S$
(source), $K$ (sink), $F$ (fork), and $C$ (combine) are computationally simple
actors without potential for GPU acceleration, and only very limited potential
for speedup through CPU-based vectorization.  On the other hand, actors $A_1,
A_2, A_3, A_4$ have GPU-accelerated versions with significant throughput gain.
In this case, however, the overall throughput gain is limited by the slowest of
the four $A_i$s so that incrementally vectorizing individual $A_i$s does not
directly impact throughput gain. 


To provide memory efficient vectorization in which this kind
of scenario is of dominant concern, we propose 
another vectorization method called {\em Mapping-Based 
Devectorization} ({\em MBD}). In contrast with ALV-based incremental
vectorization, MBD applies GLV to first vectorize all 
vectorizable actors, and
then performs devectorization on the transformed graph
derived from GLV. MBD is useful in devectorizing
actors that have relatively low CPU-based performance
gain through vectorization, and in jointly considering
vectorization improvements produced by groups of actors.


MBD performs GLV, generates a processor assignment $A$, and then evaluates for
devectorization each actor that is mapped to a CPU core in $A$. If a given
devectorization operation does not reduce the original throughput by a
pre-defined threshold $r$, the actor is devectorized.  In our experiments, we
set the threshold $r$ empirically by experimenting with different values of
$r$. We found in our experiments that $r = 0.95$ achieves the maximum
throughput gain for MBD (see Section~\ref{sec:experiments}) on the same set of 
random graphs. The optimal choice of $r$ may change for a different set of 
graphs. Alternatively, $r$ can be customized for a given graph by performing a 
search (such as a binary search) to optimize this parameter.

Although the MBD algorithm begins by applying GLV, the algorithm 
produces solutions that are in general ALV solutions. This is
because of the application of devectorization later in the algorithm,
which in general results in heterogeneous vectorization degrees
across the set of actors in the input graph.

In principle, the processor assignment $A$ can
be generated using any multiprocessor task graph scheduling technique. In our
implementation of MBD, we employ the Heterogeneous Earliest Finish Time (HEFT)
heuristic (e.g., see~\cite{augo2011x1,topc2002x1}) to generate a schedule for
the transformed graph that results from GLV, and then we extract the processor
assignment from this generated schedule.  

Devectorization saves
memory from low-impact vectorization of actors that are mapped onto
CPU cores.  When memory 
constraints are loose enough to allow GLV, the MBD technique,
based on the memory savings achieved through devectorization,
may improve throughput by allowing greater GVDs to be applied. 

Figure~\ref{fig:devect}(c) illustrates the application 
of MBD. In this example, since actors $C$, $F$, $K$, and $S$ 
are mapped onto CPU cores, they are devectorized. 
As a result of this devectorization, the buffer 
requirements on edges $(S,F)$ and $(C,K)$ are reduced to 1 for
each edge. 
Algorithm~\ref{algo:devect} provides a pseudocode description for MBD. 

\begin{algorithm}
\SetKwData{ConfigList}{$\mathit{configs}$} 
\SetKwData{Config}{$\mathit{c}$}
\SetKwData{Mapping}{$\mathit{mp}$}
\SetKwData{Flag}{$\mathit{flag}$}
\SetKwData{True}{$\mathit{true}$}
\SetKwData{False}{$\mathit{false}$}
\SetKwData{CPU}{$\mathit{CPU}$}
\SetKwData{Mapping}{$\mathit{mp}$}
\SetKwData{MappingPrime}{$\mathit{mp'}$}
\SetKwData{GVD}{$\mathit{gvd}$}
\SetKwFunction{Throughput}{throughput}
\SetKwFunction{GenMapping}{generateMapping}
\SetKwFunction{MemReq}{memSize}
\SetKwFunction{MapDevect}{mappingBasedDevectorize}
\SetKwFunction{GraphVect}{graphVectDegrees}
\SetKwFunction{Vectorize}{vectorize}
\SetKwFunction{Sort}{sortByThroughput}
\SetKwProg{Fn}{Function}{}{}

\Fn{$\MapDevect(G=(V,E),P,M)$}{
initialize $B=\{(v,1)| v \in V\}$, $\Mapping = \GenMapping(G, P)$, 
$\ConfigList = \{(B, \Mapping)\}$, $G_B = G$, $\GVD = 1$ \; 
\Repeat{$\MemReq(G_B) \leq M$}{
  $B' = B$, $\MappingPrime = \Mapping$ \; 
  $B = \GraphVect(G, \GVD)$ \; 
  $G_B = \Vectorize(G, B)$ \; 
  $\Mapping = \GenMapping(G_B)$\; 
  $\mathit{cpu\_actors} = \{v \in V | v \mathrm{\ is\ mapped\ to\ a\ CPU\ core}\}$\;
  \ForEach{$v \in \mathit{cpu\_actors}$} {
    $B'' = B - \{(v,b)\} \cup \{(v,1)\}$ \; 
    \If{$\Throughput(G,(B'',mp)) \geq r \times \Throughput(G,(B,mp)) $}{
      $B = B''$ \; 
    }
  }
  $\GVD = \GVD + 1$ \;
}
\Return $(B', \MappingPrime)$
}

\caption{Mapping-Based Devectorization (MBD).}
\label{algo:devect}
\end{algorithm}

\subsection{Mapping Actors onto HCGPs}

The $\mxsgm$-IAV and $N$-candidates IAV methods presented in
Section~\ref{sec:iav} and Section~\ref{sec:niav}, respectively, both employ
$\mxsgm$-scheduling throughout the optimization process to generate schedules
for intermediate vectorized graphs. The $\mxsgm$-scheduling approach is useful
in our iterative optimization context because it provides moderate-complexity,
bounded-buffer scheduling of multirate SDF graphs.  As mentioned in
Section~\ref{sec:iav} and Section~\ref{sec:niav}, we develop a specific
$\mxsgm$-scheduling technique called {\em Incremental Actor Re-assignment}
({\em IAR}) for use in both $\mxsgm$-IAV and $N$-candidates IAV. In this
section, we elaborate on the IAR technique.

In contrast to time-intensive scheduling methods such as Mixed Linear
Programming 
and Genetic Algorithms, IAR is designed with computational efficiency as a primary objective. This is because IAR is invoked repeatedly during each
IAV iteration --- in particular, it is invoked
for each candidate ALV configuration.

Intuitively, IAR incrementally moves actors in $\mxsgm$ schedules from
``busier'' (more loaded) processors to less busy ones.
Algorithm~\ref{algo:map} provides a pseudocode description of the IAR method.
IAR initializes the actor assignment by mapping all actors that have
GPU-accelerated versions onto the GPU, and all other actors onto a single CPU
core. This results in an initial assignment that utilizes at most two
processors (the GPU and one CPU core).

\begin{algorithm}
\SetKwInOut{Input}{input}\SetKwInOut{Output}{output}
\SetKwData{BM}{$\mathit{bestMp}$} 
\SetKwData{BTh}{$\mathit{bestTh}$}
\SetKwData{Mapping}{$\mathit{mp}$} 
\SetKwData{Th}{$\mathit{th}$}
\SetKwData{MappingPrime}{$\mathit{mp'}$} 
\SetKwData{ThPrime}{$\mathit{th'}$}
\SetKwFunction{Throughput}{throughput}
\SetKwFunction{GenMapping}{generateMapping}
\SetKwProg{Fn}{Function}{}{}
\Fn{$\GenMapping(G,P)$}{
for $v \in V$, initialize $\BM(v) = p_N$ if $t(v,p_N) < \infty$ and $\BM(v) = p_1$ otherwise \; 
initialize $\BTh = \Throughput(G,\BM)$ \; 
\ForEach{$v \in V$}{
  $\Mapping = \BM$, $\Th = \BTh$, $p^* = \BM(v)$ \; 
  $Q = \{q \in P | q \neq p^*\}$ \;
  \ForEach{$p \in Q$}{
    $\MappingPrime = \Mapping - \{(v, \Mapping(v))\} \cup \{(v,p)\}$ \; 
    $\ThPrime = \Throughput(G, \MappingPrime)$ \; 
    \lIf{$\ThPrime > \Th$}{
      $\Mapping = \MappingPrime$, $\Th = \ThPrime$ 
    }
  }
  \lIf{$\Th > \BTh$}{
    $\BM = \Mapping$,$\BTh = \Th$
  }
}
\Return $(\BM, \BTh)$
}
\caption{Incremental Actor Re-assignment (IAR).}
\label{algo:map}
\end{algorithm}


Then IAR iteratively computes the maximum throughput gain for all
actor-processor pairs, and selects the pair that gives the highest throughput
at each iteration. In this context, selection of an actor-processor pair $(a,
p)$ means that the current processor assignment of actor $a$ will be discarded,
and actor $a$ will be assigned (``moved'') to processor $p$. For
this selection process, only actors that have not yet been selected during
previous iterations are considered. The throughput gain is computed with the
aid of the function denoted in Algorithm~\ref{algo:map} as \verb#throughput#.
This function invokes the simulation-based throughput estimator discussed in
Section~\ref{sec:throughput}. Each actor is moved only once during
execution of IAR. 


\subsection{Throughput Estimation}
\label{sec:throughput}

For compile-time throughput estimation, we have developed a throughput
simulator for SDF graphs that follows bounded-buffer execution semantics
(defined in Section~\ref{sec:background}) with a
statically-determined processor assignment, as derived by the $\mxsgm$ strategy
introduced in Section~\ref{sec:defs}. The inputs to the simulator are: (1) the
transformed SDF graph $G_v$ that results from the candidate set of
vectorization operations that is under evaluation; (2) the $\mxsgm$ mapping for
$G_v$ that is generated by IAR; (3) the $\Omega$ buffer bound for each edge in
$G_v$; (4) an estimate of the execution time for each actor in $G_v$; and (5)
an estimate of the context transfer time between the main memory and the device
memory on the target platform.  

To estimate the throughput of a vectorized SDF graph, we first map vectorized
actors onto processors, and follow the approach of $\mxsgm$-scheduling defined
in Definition~\ref{def:omega} to compute the schedule. Throughput is then
estimated by simulating the execution of the derived schedule.
In our experiments, the execution time
estimates under different vectorization degrees for each actor as well as the
context transfer time are derived by using measurements of actor and context 
transfer execution on the target HCGP. 

\subsection{Summary}

\begin{figure}[]
\centering
\includegraphics[width=2.7in]{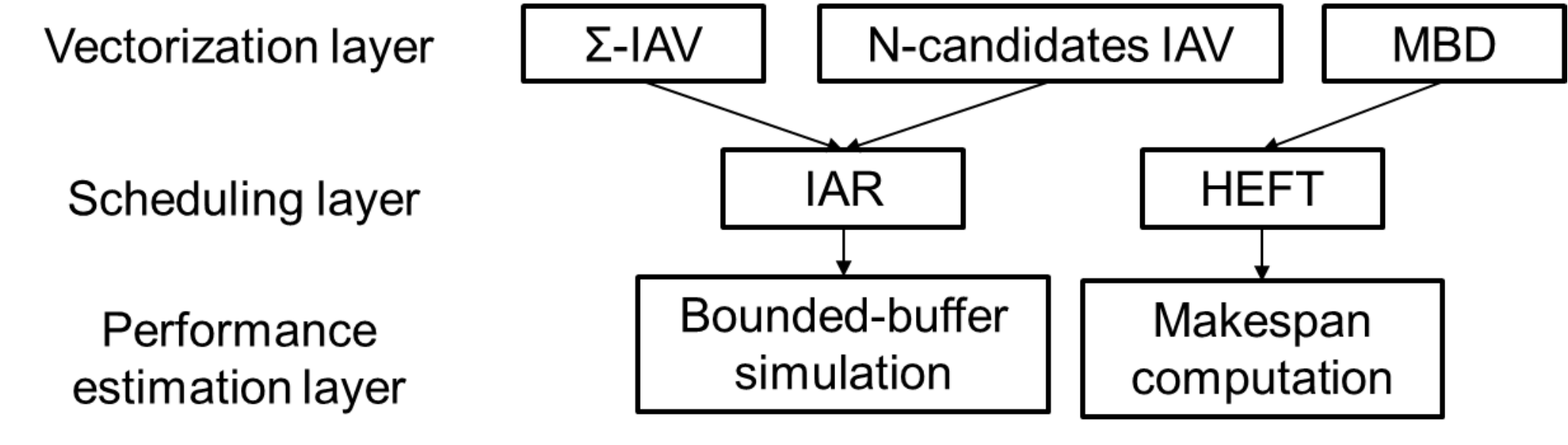}
\caption{Layered structure of vectorization, scheduling, and performance 
estimation in the proposed design optimization framework.}
\label{fig:layers}
\end{figure}

Figure~\ref{fig:layers} summarizes the developments of this section by
illustrating relationships among the key analysis and optimization techniques 
that have been introduced. Recall that IAV, HEFT, and MBD stand,
respectively, for incremental actor vectorization, heterogeneous
earliest finish time, and mapping-based devectorization. Each directed edge in 
Figure~\ref{fig:layers}
represents usage of one technique (at the sink of the edge)
by another (at the source of the edge). For example, IAR is
used by $\mxsgm$-IAV. 

\section{Experiments using Synthetic Graphs}
\label{sec:experiments}

In this section, we demonstrate the effectiveness of 
the models and methods developed in Section~\ref{sec:vect-sched} through
experiments that study throughput gain and running time.
We compare our methods with the approach
of applying graph-level vectorization (GLV) followed by
task-graph scheduling. 

We use Heterogeneous Earliest Finish Time (HEFT) as the task-graph scheduling
method in this comparison. HEFT is a commonly used task-graph scheduling method
for HCGPs (e.g., see~\cite{augo2011x1}). The integration of HEFT with GLV can
be viewed as a natural way to integrate SDF vectorization and scheduling using
conventional techniques. We refer to the combination of GLV and HEFT as the
{\em GLV-HEFT baseline}  or simply as {\em GLV-HEFT}. As implied by this
terminology, GLV-HEFT is employed in this experimental study as a baseline for
evaluating our proposed methods. The GLV-HEFT baseline applies both GPU
acceleration and CPU-GPU multi-processor scheduling.  We demonstrate in this
section that the ALV and IAR scheduling methods developed in this paper provide
significant throughput gain over this baseline approach under given memory
constraints. 

\subsection{Experimental Setup}

We have developed an integrated software synthesis framework called {\em
DIF-GPU} to provide a streamlined workflow that combines actor-level /
graph-level vectorization, multi-rate / single-rate SDF scheduling, code
generation, and runtime support on heterogeneous computing platforms with
multi-core CPUs and GPUs. For details about the DIF-GPU framework, we refer
readers to~\cite{lin2016x1}. 

In the experiments presented in this section, we 
employ an HCGP consisting of a quad-core Intel 
i5-6400 CPU and an NVIDIA Geforce GTX750 GPU. Actor implementations that
are developed for multi-core CPU and GPU execution
are compiled using GCC 4.6.3 and the 
NVIDIA CUDA compiler (NVCC) 7.0, respectively. 

\subsection{Synthetic Graph Generation}
\label{sec:synthetic}

We use Task Graphs For Free (TGFF)~\cite{dick1998x2} to generate large sets of
synthetic SDF graphs with varied size and complexity. Key parameter settings
that we use in TGFF are as follows: the maximum in-degree and out-degree for
graph nodes are both set to 3, and the average and multiplier for the lower
bound on the number of graph nodes are both set to 20. 

From the graph
topologies generated by TGFF, we randomly map each graph vertex to a specific
DSP actor type that has both a CPU-targeted and GPU-targeted implementation. We
perform this vertex-to-actor mapping for all actors in each randomly-generated
graph. A broad set of DSP actor types --- including actors for
cross-correlation, FIR filtering, FFT computation, and vector algebra --- are
considered when performing this mapping. The GPU-accelerated implementations of
these actors provide speedups from 1X to 20X compared to the corresponding
multicore CPU implementations. This use of TGFF in conjunction with randomly
generated actor mappings helps us to evaluate the performance of our proposed
methods on a large variety of graph topologies.


In our experiments, the source and sink actors are selected from a pool of
different implementations of data sources and sinks. Because the input/output
interfacing functionality in an embedded HCGP is typically implemented on a
CPU, we assume that source and sink actors can only be mapped onto CPU cores.

We profile the actors by measuring the execution times of the actors' firings
on the target platform under a series of vectorization degrees. This profiled
data is then used as input to the evaluated vectorization and scheduling
techniques. The profiled data is also used to simulate the
vectorization-integrated schedules that are derived from the proposed and
baseline techniques. This simulation is based
on the throughput simulator presented in Section~\ref{sec:throughput}. We use simulation here to enable efficient, automated
comparisons across a large variety of different graph structures. In
Section~\ref{sec:ofdm}, we complement this simulation-based evaluation approach
with our experimental evaluation of a case study involving an orthogonal
frequency-division multiplexing (OFDM) receiver. The evaluation in
Section~\ref{sec:ofdm} is performed by synthesizing software using DIF-GPU for
the targeted HCGP platform, executing the synthesized software on the target
platform, and measuring the resulting execution time performance.


\subsection{Vectorization} 
\label{sec:vect-res}

In this section, we apply the different ALV methods introduced in
Section~\ref{sec:vect-sched} to a large collection of synthetic SDF graphs, and
evaluate the performance of the derived schedules by simulating their
bounded-buffer execution.  Note that the baseline for speedup here is GLV-HEFT,
where extensive vectorization has been applied, and both the CPU and GPU are
used to schedule dataflow actors. This is a much ``higher'' baseline than
single- / multi-core CPU implementation without vectorization.  Therefore, the
speedup computed over this baseline is relatively small.  The synthetic graphs
are generated using TGFF together with randomized vertex-to-actor mappings, as
described in Section~\ref{sec:synthetic}.  We evaluate the speedup over the
GLV-HEFT baseline under different memory constraints. 

To compare speedups across SDF graphs that have different sizes (i.e.,
different numbers of actors and edges) and different multirate properties (as
defined by the production and consumption rates on the actor ports), we
introduce a concept of {\em relative memory bounds} as a normalized
representation for memory constraints. Given an algorithm $A$ for performing
GLV, the relative memory bound $M(G)$ for an SDF graph $G$ is defined as $M(G)
= M_0 \times \alpha$, where $M_0$ is the memory cost of the GLV solution
derived by Algorithm $A$ when applied to $G$ with $\mathrm{GVD} = 1$, and
$\alpha$ is a constant that represents the ``tightness/looseness'' of the
applied memory constraint. We experiment with $\alpha \in
\{1.0, 1.5, \ldots, 4.5, 5\}$ to cover a series of memory constraints ranging,
respectively, from tight to loose. 

\begin{figure}[]
\centering
\includegraphics[width=3in]{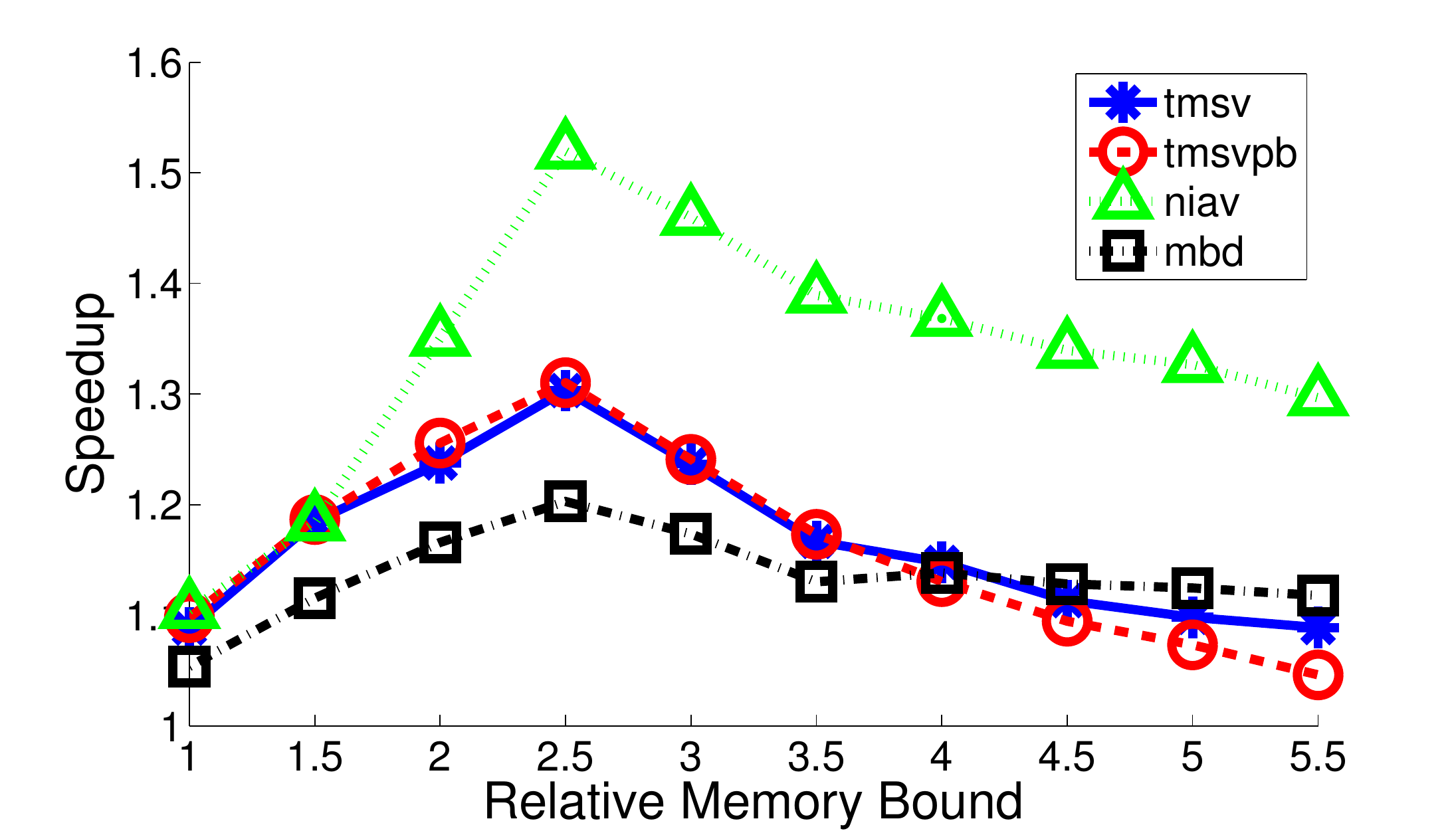}
\caption{Average speedups measured for four ALV techniques 
introduced in Section~\ref{sec:vect-sched}: TMSV $\mxsgm$-IAV, 
TMSVPB $\mxsgm$-IAV, $N$-candidates IAV, and MBD. }
\label{fig:sim-alv}
\end{figure}

Figure~\ref{fig:sim-alv} shows the average simulated speedup 
that we measured from a set of randomly generated SDF graphs for different 
techniques for ALV that were introduced in Section~\ref{sec:vect-sched}. 
As mentioned previously, these speedups are in comparison to baseline 
solutions that are derived using the GLV-HEFT baseline technique.  
These results are for a target
platform configuration that consists of 1 CPU core and 1 GPU. Here, ``TMSV
$\mxsgm$-IAV'' and ``TMSVPB $\mxsgm$-IAV'' represent the $\mxsgm$-IAV algorithm
with the TMSV and TMSVPB score functions, respectively.  The measured
throughput gain ranges from 0.8X to 2.4X, and also exhibits significant variation
from one SDF graph to another. 

We refer to {\em ALV-IAR} as the
meta-algorithm that results from applying all four of the proposed ALV
techniques, and selecting the best result from among the four derived
solutions. In Section~\ref{sec:ofdm}, we perform further experimental analysis
of the ALV-IAR method, which provides a way to leverage complementary benefits
of all of the key ALV techniques introduced in Section~\ref{sec:vect-sched}.
ALV-IAR is useful, in particular, for design scenarios that can tolerate the
relatively large optimization time that is required by $N$-candidates IAV, which
dominates the time required by ALV-IAR. ALV-IAR
demonstrates average and maximum speedup values of 1.36X and 2.9X on the
benchmark set. 

We see that $N$-candidates IAV provides the
largest average speedup by a significant margin, and this algorithm also
provides the largest maximum speedup.  We anticipate that this is because
$N$-candidates IAV uses more vectorization candidate solutions throughout the
search process.  The other three ALV techniques achieve similar average and
maximum throughput gain. 

We have also observed that the average speedup of ALV methods increases until
$M(G) = 2.5$ and then gradually drops off. When $M(G)$ is close to 1, there is
little room for vectorization, so ALV and GLV achieve similar throughput, and
the average speedup is close to 1. As $M(G)$ increases to 2.5, more flexibility
is provided for ALV to vectorize for better performance than GLV.  When $M(G)$
increases beyond 2.5, the memory is sufficient to allow relatively large
vectorizations for all actors, so the throughput gain from enabling further
vectorization is worn off. 

Although the MBD method and the two $\mxsgm$-IAV methods achieve smaller
average speedup compared to \mxniav{}, they run significantly faster
(see Section~\ref{sec:runtime}), and can be
useful in cases where quicker turnaround time is desired from the software
synthesis process. In addition, there are cases where they perform better than
\mxniav{}.

\subsection{Runtime} 
\label{sec:runtime}

In this section, we compare the measured running times of the four proposed ALV
techniques.  We tested the running times of the ALV techniques on the same set
$S_g$ of randomly generated SDF graphs that we used in the experiments
reported on in Section~\ref{sec:vect-res}.  The set
$S_g$ consists of $120$ graphs, where the of number of nodes in a given graph
ranges from $3$ to $30$.

\begin{figure}[]
\centering
\subfigure[]{
  \includegraphics[width=2.6in]{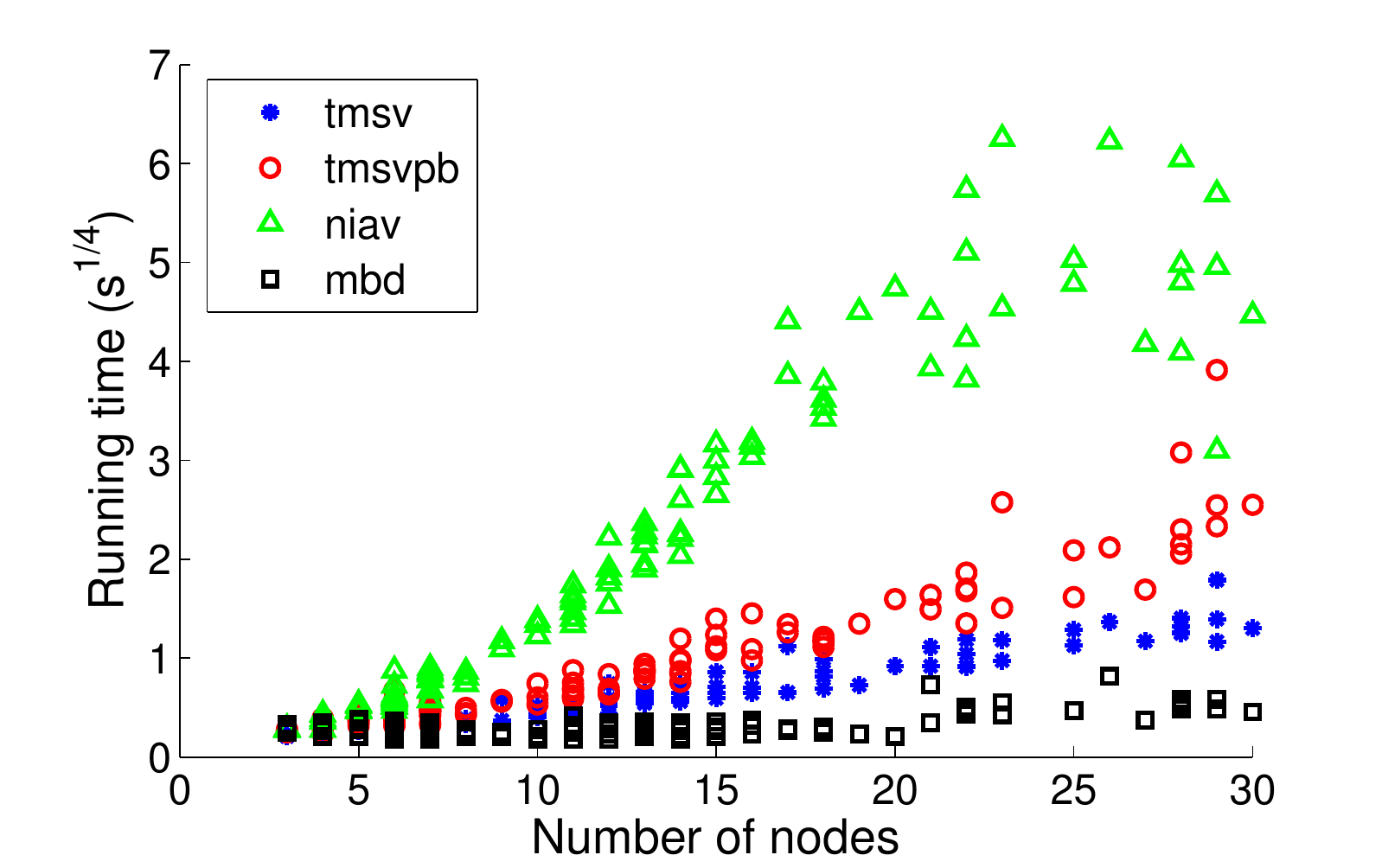}
  \label{fig:rt-a}
}
\subfigure[]{
  \includegraphics[width=2.6in]{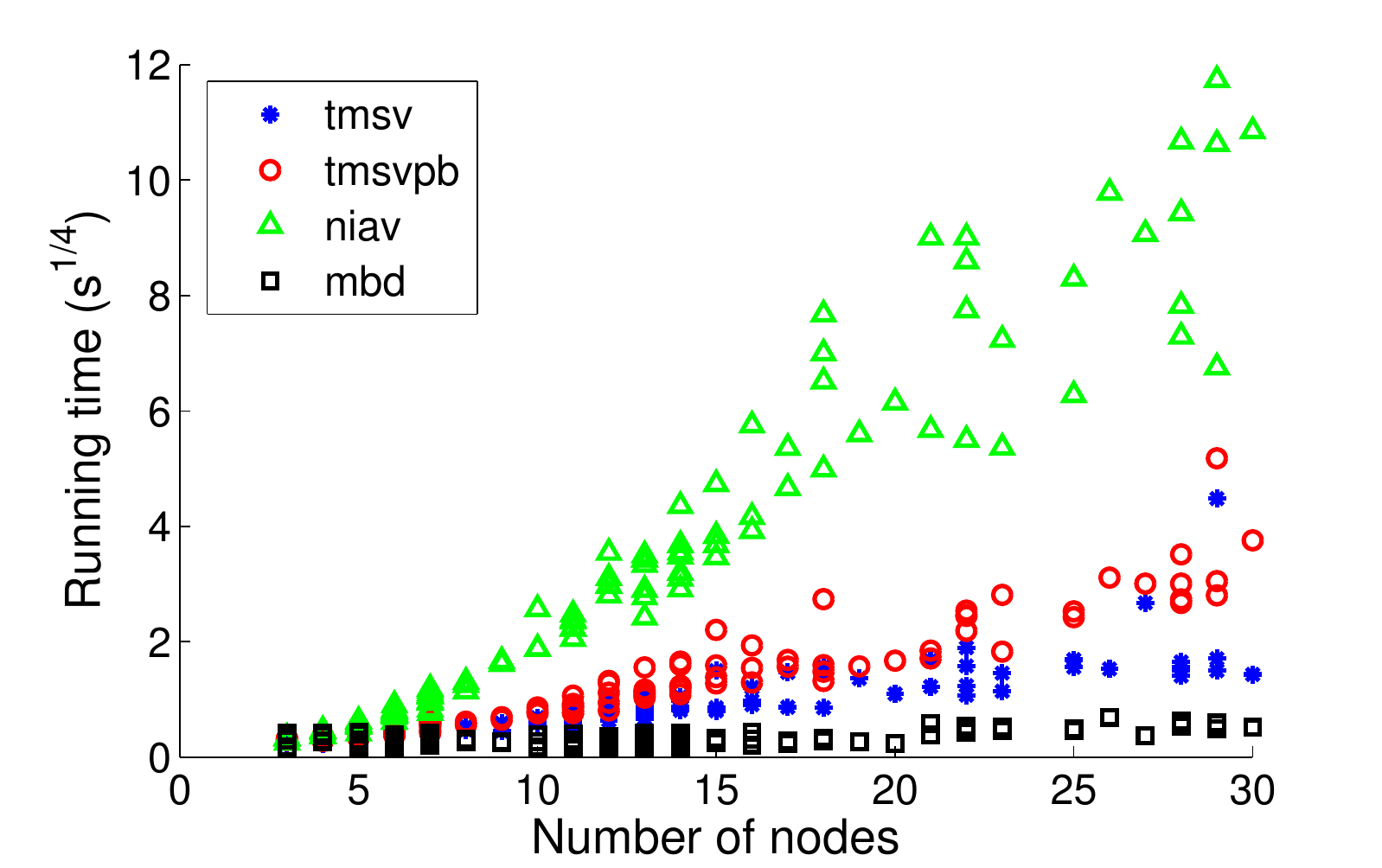}
  \label{fig:rt-b}
}
\caption{Runtime of ALV methods under different memory constraints: 
(a) $M=2M_0$, and (b) $M=4M_0$. }
\label{fig:rt}
\end{figure}

Figure~\ref{fig:rt} shows the measured running times for the four ALV methods
with respect to the number of nodes in the input graph. For each of the four
ALV methods, there are 120 points plotted in each part of the figure --- one
point for each graph in $S_g$.  Thus, Figure~\ref{fig:rt}(a) and
Figure~\ref{fig:rt}(b) each depicts a total of $4 \times 120 = 480$ plotted
points. 

Figure~\ref{fig:rt} presents running time results associated with two different
memory constraints --- $M = 2M_0$ in Figure~\ref{fig:rt}(a), and $M = 4M_0$ in
Figure~\ref{fig:rt}(b) (see the discussion on relative memory bounds in
Section~\ref{sec:vect-res}). These two memory constraints are used to represent
relatively tight and loose memory budgets, respectively. The vertical axes in
Figure~\ref{fig:rt} correspond to $s^{1/4}$, where $s$ is the measured running 
time in seconds. Here, we apply an exponent of $(1/4)$ to help improve
clarity in depicting the large number of displayed points.

The list of the ALV methods sorted from the fastest to the slowest are: MBD,
$\mxsgm$-IAV with the TMSV score function, $\mxsgm$-IAV with the TMSVPB score
function, and NIAV. Note that the TMSVPB score function runs more slowly
compared to TMSV due to the computation cost of $\mxomegabuf(G_{B'})$ in the
denominator of Equation~\ref{eq:tmsvpb}.  This cost involves recomputing the
buffer requirements for all of the edges in $G$.  Table~\ref{tab:rt} shows the
running times of the ALV methods on a specific graph with 22 nodes and 33
edges.  This graph is selected randomly to provide further insight into
variations in running time among the four ALV methods. 

In our experiments, we find that typically MBD finishes within 1 second, while
the running times of the two $\mxsgm$-IAV methods usually range from several
seconds up to a few minutes. We expect that this kind of running time profile
is acceptable in many coarse grain dataflow design scenarios in the embedded
signal processing domain, where actors typically perform higher level signal
processing operations, and therefore the number of nodes in the graphs is
limited compared to other types of dataflow graphs that are based on
fine-grained actors.

\begin{table}
\caption{The running times (in seconds) of the ALV methods on 
a specific SDF graph with 22 nodes and 33 edges. 
\label{tab:rt}}{%
\begin{tabular}{|c|c|c|c|c|}
\hline
  & TMSV $\mxsgm$-IAV & TMSVPB $\mxsgm$-IAV & NIAV & MBD \\
\hline
$M=2M_0$ & 2.0 & 8.4 & 320 & 0.1\\
\hline
$M=4M_0$ & 13.0 & 35.9 & 3500 & 0.7\\
\hline
\end{tabular}}
\end{table}

The running time of \mxniav{} is generally the longest among all four methods,
and grows rapidly with the number of nodes. In our experiments with an SDF
graph having 30 nodes, for example, \mxniav{} takes 3 hours to finish its
computation.  Therefore, \mxniav{} is more suitable in situations when the SDF
graph is relatively small, design turnaround time is not critical, or solution
quality is of utmost importance.

\section{Case Study: OFDM}
\label{sec:ofdm}

\begin{figure}[]
\centering
\includegraphics[width=4in]{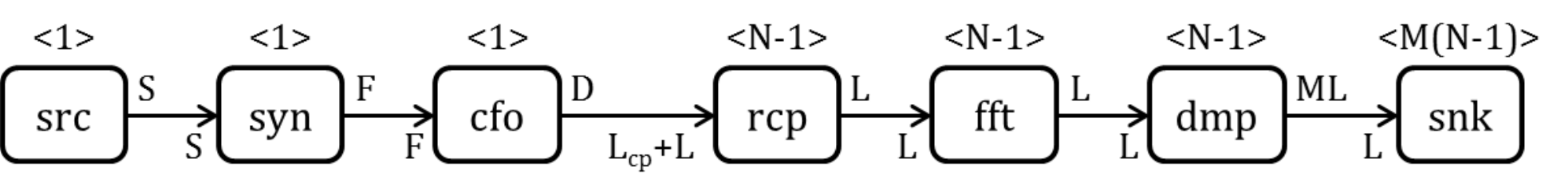}
\caption{SDF model of OFDM-RX application. }
\label{fig:ofdm-rx}
\end{figure}

In this section, we demonstrate the effectiveness of our new
ALV-integrated software synthesis framework through a case study involving an
{\em orthogonal frequency-division multiplexing} ({\em OFDM}) receiver ({\em
OFDM-RX}). The OFDM-RX is an adapted version of the OFDM system described
in~\cite{mass2012x1}. Figure~\ref{fig:ofdm-rx} shows an SDF model for the
OFDM-RX application. The value above each actor in Figure~\ref{fig:ofdm-rx}
gives the repetition count of the actor. Table~\ref{tab:ofdm-actors} lists the
actors in this SDF model and describes their corresponding functions.  The
system can operate with different parameter values, as shown in
Table~\ref{tab:ofdm-params}. 


\begin{table}
\caption{Actors in the OFDM-RX application.\label{tab:ofdm-actors}}{%
\begin{tabular}{| l | l | }
\hline
Actor & Description \\
\hline
$\mxsrc$ & Read samples of the input signal. \\
\hline
$\mxsyn$ & Perform time-domain synchronization. \\
\hline
$\mxcfo$ & Remove carrier frequency offsets. \\
\hline
$\mxrcp$ & Remove cyclic prefix. \\
\hline
$\mxfft$ & Perform Fast-Fourier Transform on symbols.\\
\hline
$\mxdmp$ & Map OFDM symbols into bit stream.\\
\hline
$\mxsnk$ & Write bit stream onto the output.\\
\hline
\end{tabular}
}
\end{table}

\begin{table}
\caption{Parameters in the OFDM-RX application model, along with the
settings or values we use in our
experiments. \label{tab:ofdm-params}}{%
\begin{tabular}{| m{0.3cm} | m{4cm} | m{2.8cm} |}
\hline
 & Description & Values\\
\hline
$L$ & Number of subcarriers per OFDM symbol & [128, 256, 512, 1024] \\
\hline
$N$ & Number of OFDM symbols per frame & 10 \\
\hline
$L_{\mxcp}$ & Length of cyclic prefix for each OFDM symbol & $(9/128)L$ \\
\hline
$M$ & Number of bits per sample & 4 \\
\hline
$D$ & Length of data excluding training symbols& $(N-1)(L+L_{\mxcp})$ \\
\hline
$F$ & Length of a frame & $N(L+L_{\mxcp})$ \\
\hline
$S$ & Size of sample stream & $2F$ \\
\hline
\end{tabular}
}
\end{table}

\subsection{System Implementation and Profiling}

We have implemented the OFDM-RX actors using the {\em Lightweight Dataflow
Environment} ({\em LIDE}), which provides a programming methodology and
associated application programming interfaces (APIs) for implementing dataflow
graph actors and edges in a wide variety of platform-oriented languages, such
as C, C++, CUDA, and Verilog~\cite{shen2010x4,shen2011x4}.  In our OFDM-RX
system, GPU-accelerated implementations are available for all actors other than
the $\mxsrc$ and $\mxsnk$ actors. The $\mxsrc$ and $\mxsnk$ actors are not
mapped to the GPU in our implementation because of input/output operations that
are involved in these actors.

\begin{figure}[]
\centering
\subfigure[]{
  \includegraphics[trim={3.2cm 0.1cm 0.1cm 1cm},clip=true, width=1.8in]
  {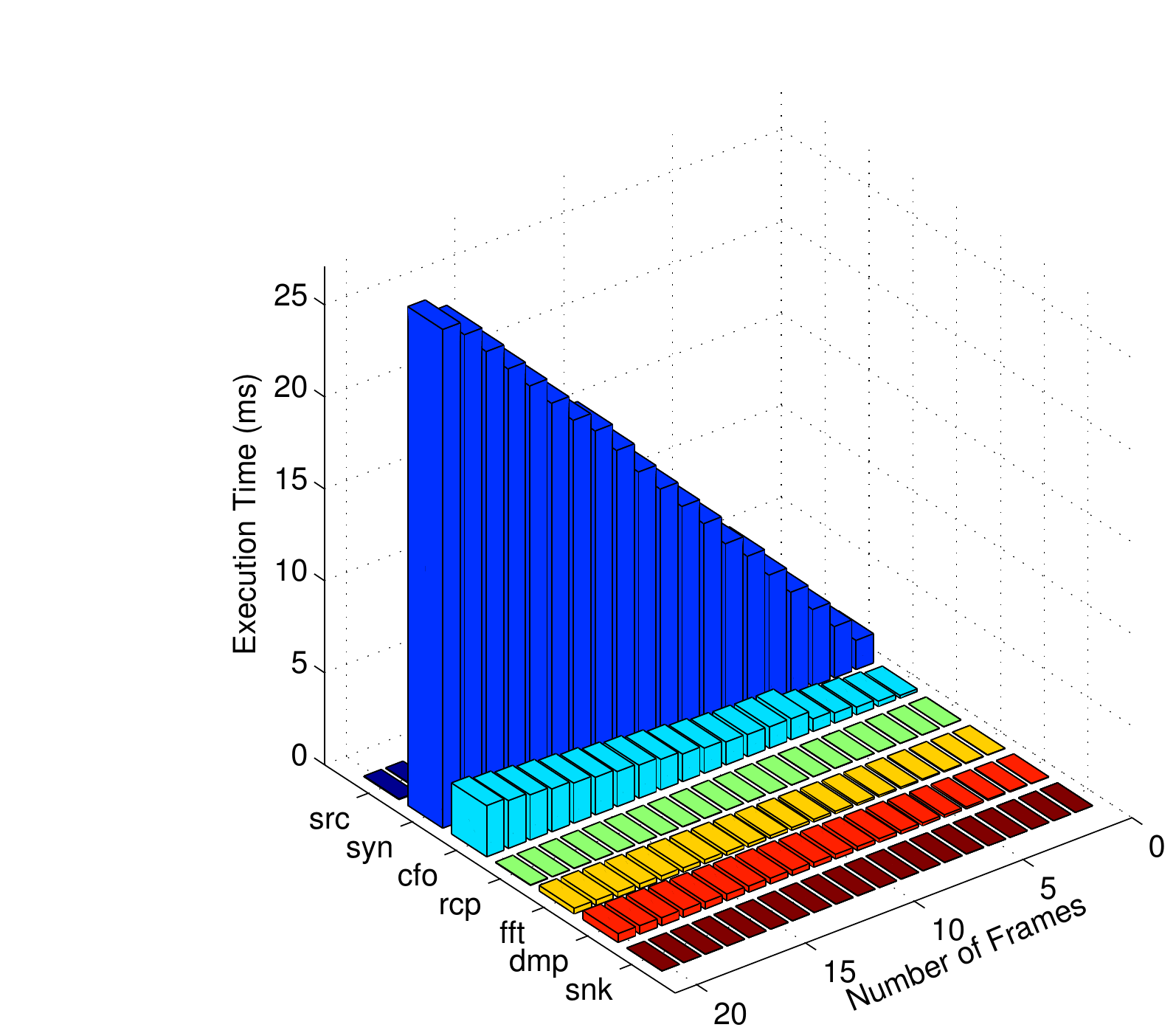}
  \label{fig:actor-cpu-times}
}
\subfigure[]{
  \includegraphics[trim={3.1cm 0.1cm 0.1cm 1cm},clip=true, width=1.8in]
  {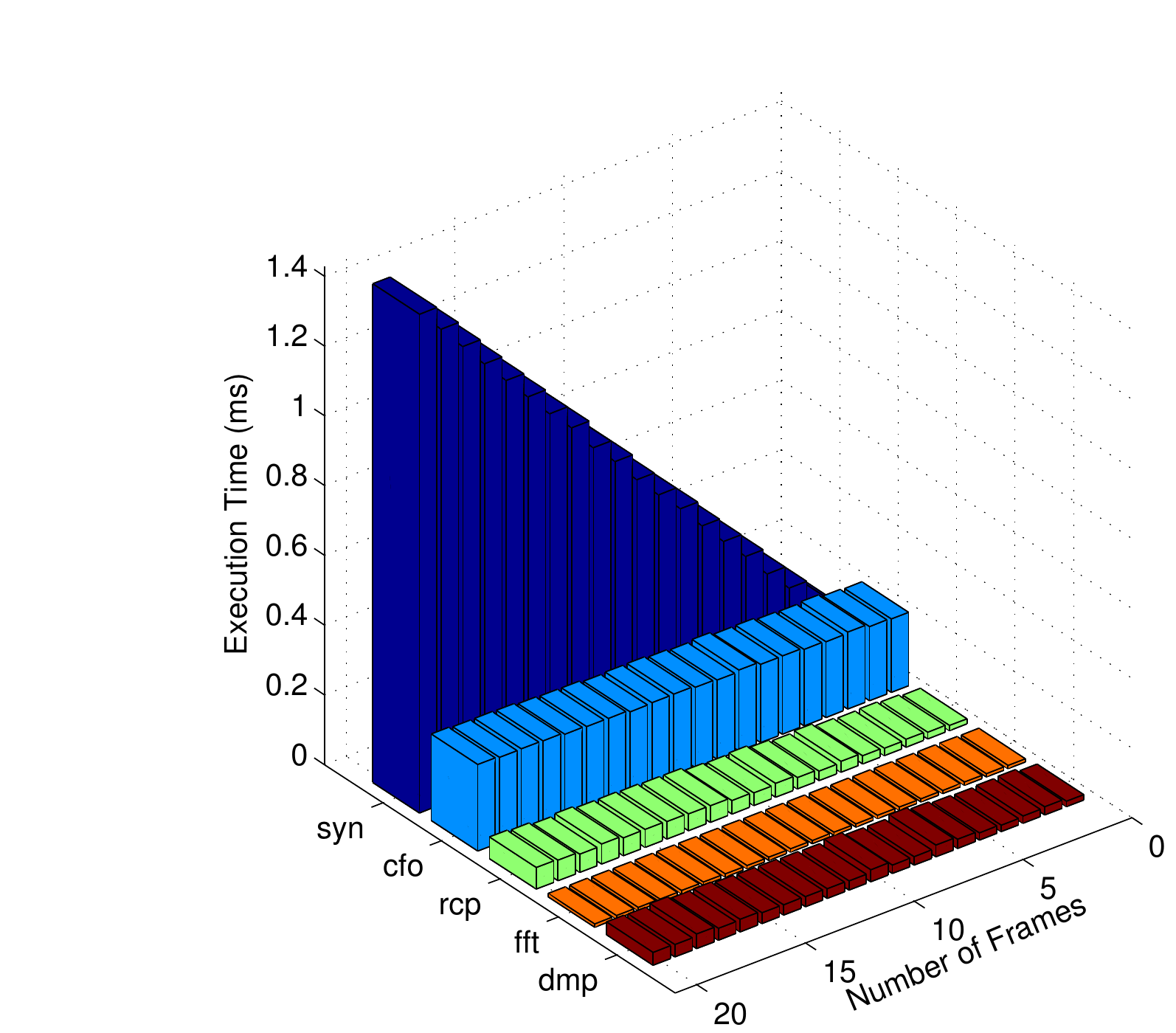}
  \label{fig:actor-gpu-times}
}
\caption{$t_T(v)$s on (a) CPU and (b) GPU for OFDM-RX actors that are vectorized to 
process multiple frames in each firing.}
\label{fig:actor-times}
\end{figure}

We have profiled the execution times for the OFDM-RX actors on both the CPU and
GPU.  Figure~\ref{fig:actor-times} summarizes the average execution times per
SDF graph iteration for the actors. This average time can be expressed as
$t_T(v) = q(v)t(v)$, where $q$ represents the repetitions vector of the
enclosing SDF graph, and $t(v)$ represents the average execution time measured
for a single firing of $v$.  These execution time estimates are measured on
both the CPU and GPU when $L=256$, and the actors are vectorized to process
different numbers of data frames per vectorized invocation.  Observe from
Figure~\ref{fig:actor-times} that the distribution of the $t_T(v)$ in OFDM-RX
are uneven, and that the $\mxsyn$ and $\mxcfo$ actors dominate the execution
times both on the CPU and GPU.  Also, observe that although actor execution
times are roughly proportional to the number of frames $N_F$, they increase at
different rates in relation to $N_F$ --- for example, $t_T(\mxcfo)$ on the GPU
grows very slowly with increases in $N_F$, and $t_T(\mxsyn)$ grows much faster. 


\subsection{Software Synthesis with GLV-HEFT}

We first measure the performance improvement achieved by GLV-HEFT when
integrated in our DIF-GPU software synthesis framework.  Here, we measure the
system throughput under 11 different configurations without any memory
constraints imposed.  These measurements are performed on software
implementations that are generated automatically using DIF-GPU integrated with
GLV-HEFT. 

In contrast to the relative throughput metric (see Section~\ref{sec:defs}) that
is used as a general performance metric in Section~\ref{sec:experiments}, we 
employ {\em frames per second} as the throughput metric
more specific to the OFDM-RX application.  

We denote the results (throughput values) from these measurements by
$\mxth_0, \mxth_1, \ldots, \mxth_{10}$.  Here, $\mxth_0$, denotes the
throughput when the input graph is not vectorized and all actors are mapped
onto a single CPU core.  On the other hand, for $b \in \{1, 2, \ldots, 10\}$,
$\mxth_b$ represents the throughput obtained when GLV is applied with 
$\mathrm{GVD} = b$, and HEFT is used to schedule the resulting vectorized graph
(GLV-HEFT)~\cite{lin2016x1}.  

\begin{figure}[]
\centering
\includegraphics[width=4in]{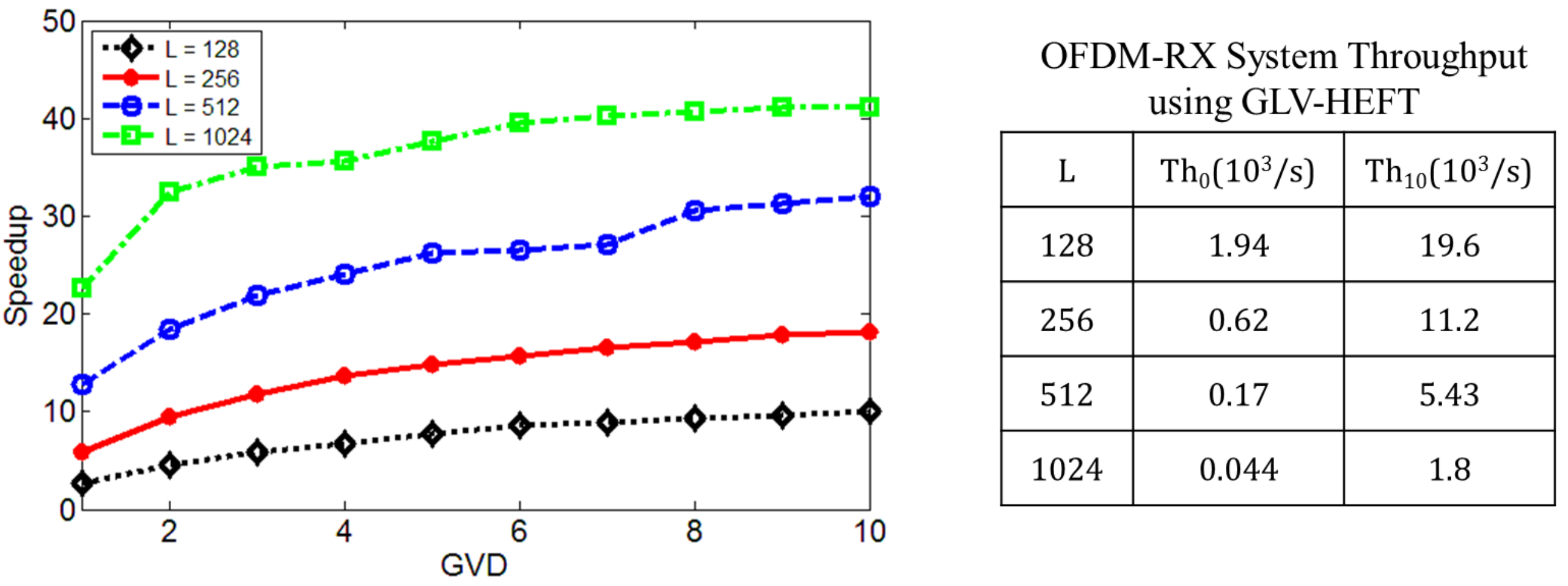}
\caption{Speedup of the OFDM-RX application over a single CPU implementation 
for different GVD values and different values of the bandwidth parameter
$L$.}
\label{fig:sp-unbound}
\end{figure}

Figure~\ref{fig:sp-unbound} shows the speedup in throughput 
of GLV over the single-CPU implementation, and compares $\mxth_0$ and 
$\mxth_{10}$ in more detail for different values of $L$. The maximum measured
speedups achieved here are 10.1X, 18.1X, 31.9X, 41.1X for 
$L=128,256,512,1024$, respectively. 

In our experiments, actors in the dataflow model are coarse-grained
signal-processing modules. Before vectorization, the actors already encapsulate
multiple steps of processing on large signal arrays, and extensively utilize
GPU data-parallelism. For example, the unvectorized syn actor in the OFDM
receiver application consists of multiple steps of cross-correlation on 20 OFDM
symbols. Therefore, saturation at small GVD levels can be expected in
Figure~\ref{fig:sp-unbound}. 



\subsection{Software Synthesis with ALV-IAR}

In this section, we perform measurements and comparisons that involve software
implementations that are generated automatically using DIF-GPU integrated with
ALV-IAR.  The experiments are
performed under different memory budgets and different levels of bandwidth $L$
(an application-level parameter).  For comparison, we apply DIF-GPU integrated
with GLV-HEFT to synthesize software that incorporates vectorized schedules
constructed using GLV-HEFT instead of ALV-IAR.

Table~\ref{tab:conf} shows an example of the vectorization degrees and
processor assignments derived for OFDM-RX under a specific memory constraint.
This memory constraint is selected to represent one that is neither very tight
nor very loose.  These vectorized scheduling results are derived by ALV-IAR,
and the throughput is measured by executing the resulting software
implementation that is synthesized by DIF-GPU. The vectorization and processor assignment (mapping)
results are shown in Table~\ref{tab:conf} as lists of values that correspond to
the graph actors in their topological order ($\mxsrc$, $\mxsyn$, \ldots,
$\mxsnk$). The numbers $0$ and $1$ in the Mapping column represent the CPU-core
and GPU, respectively.  The results in Table~\ref{tab:conf} show that ALV-IAR
produces a 1.2X speedup compared to the baseline technique for the selected
memory constraint.

\begin{figure}[]
\centering
\subfigure[]{
  \includegraphics[width=2in]
  {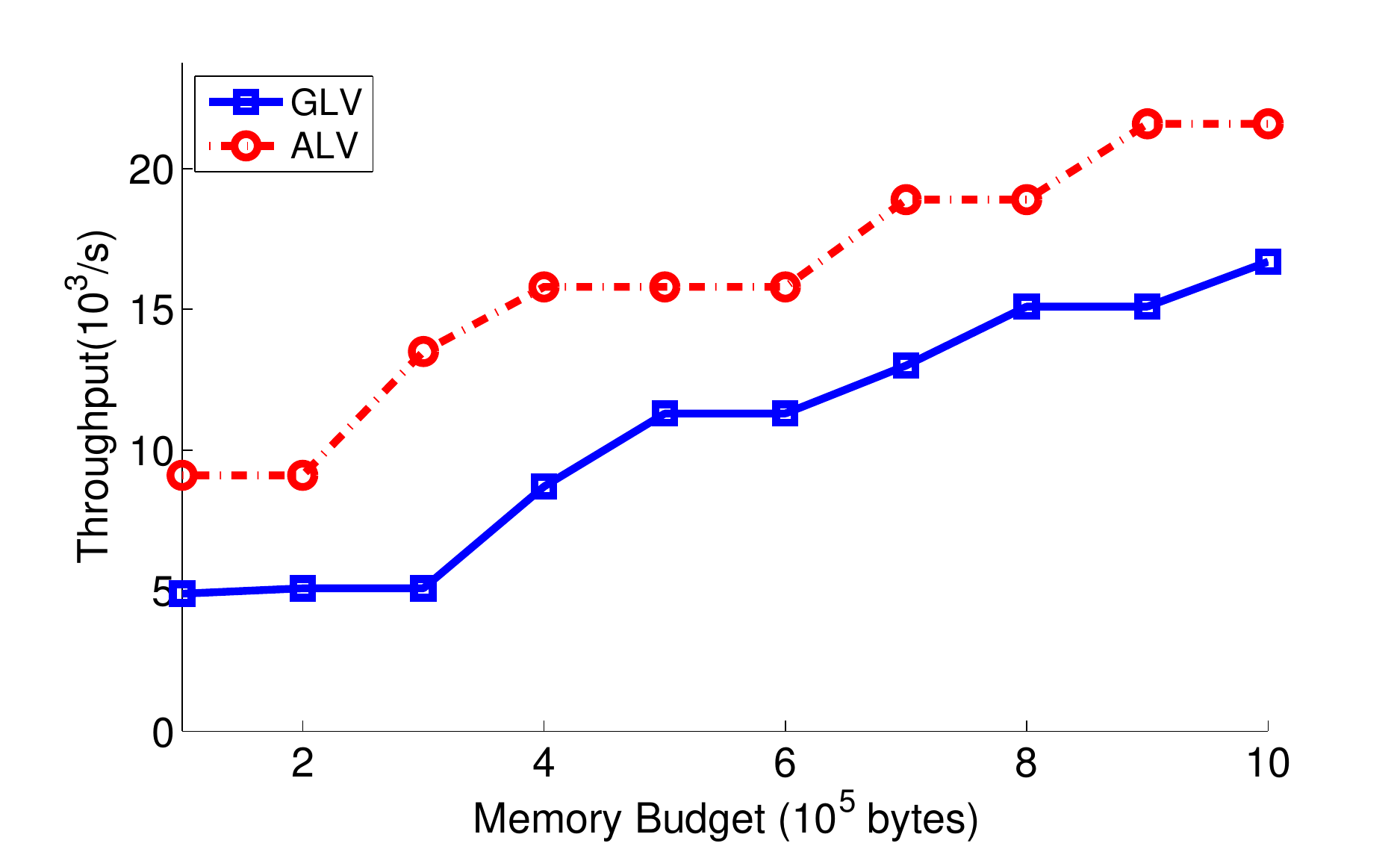}
  \label{fig:vect-128}
}
\subfigure[]{
  \includegraphics[width=2in]
  {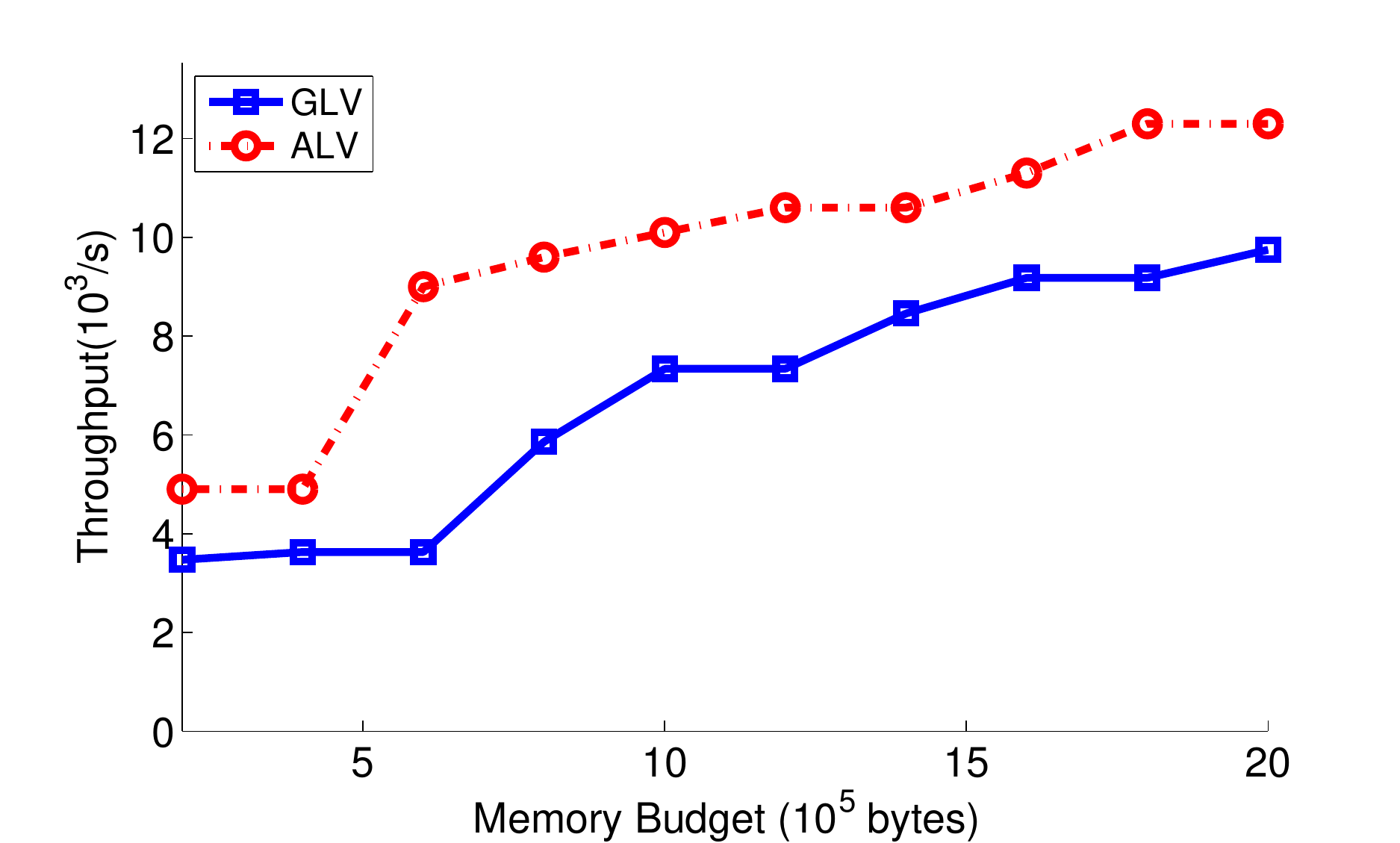}
  \label{fig:vect-256}
}
\subfigure[]{
  \includegraphics[width=2in]
  {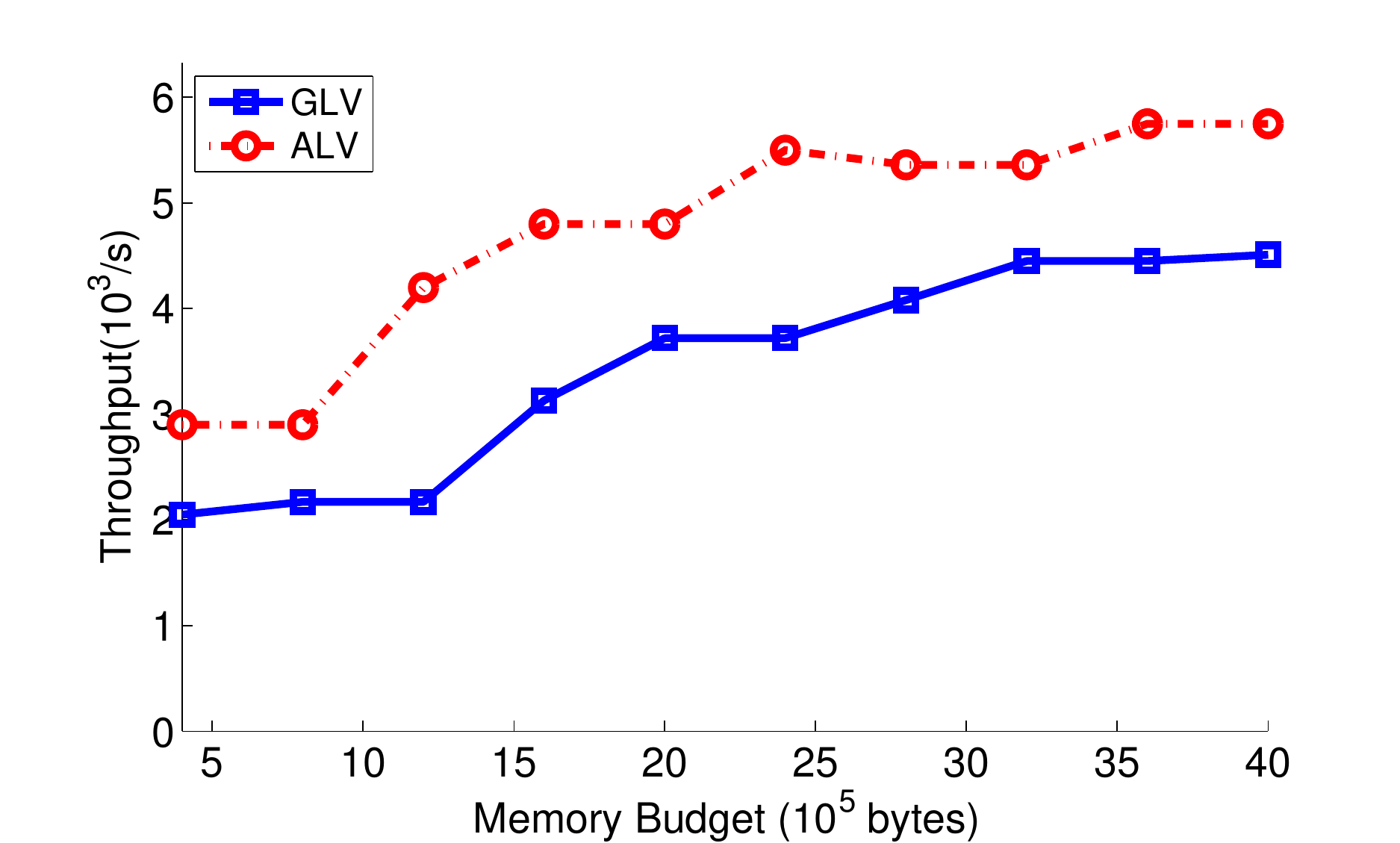}
  \label{fig:vect-512}
}
\subfigure[]{
  \includegraphics[width=2in]
  {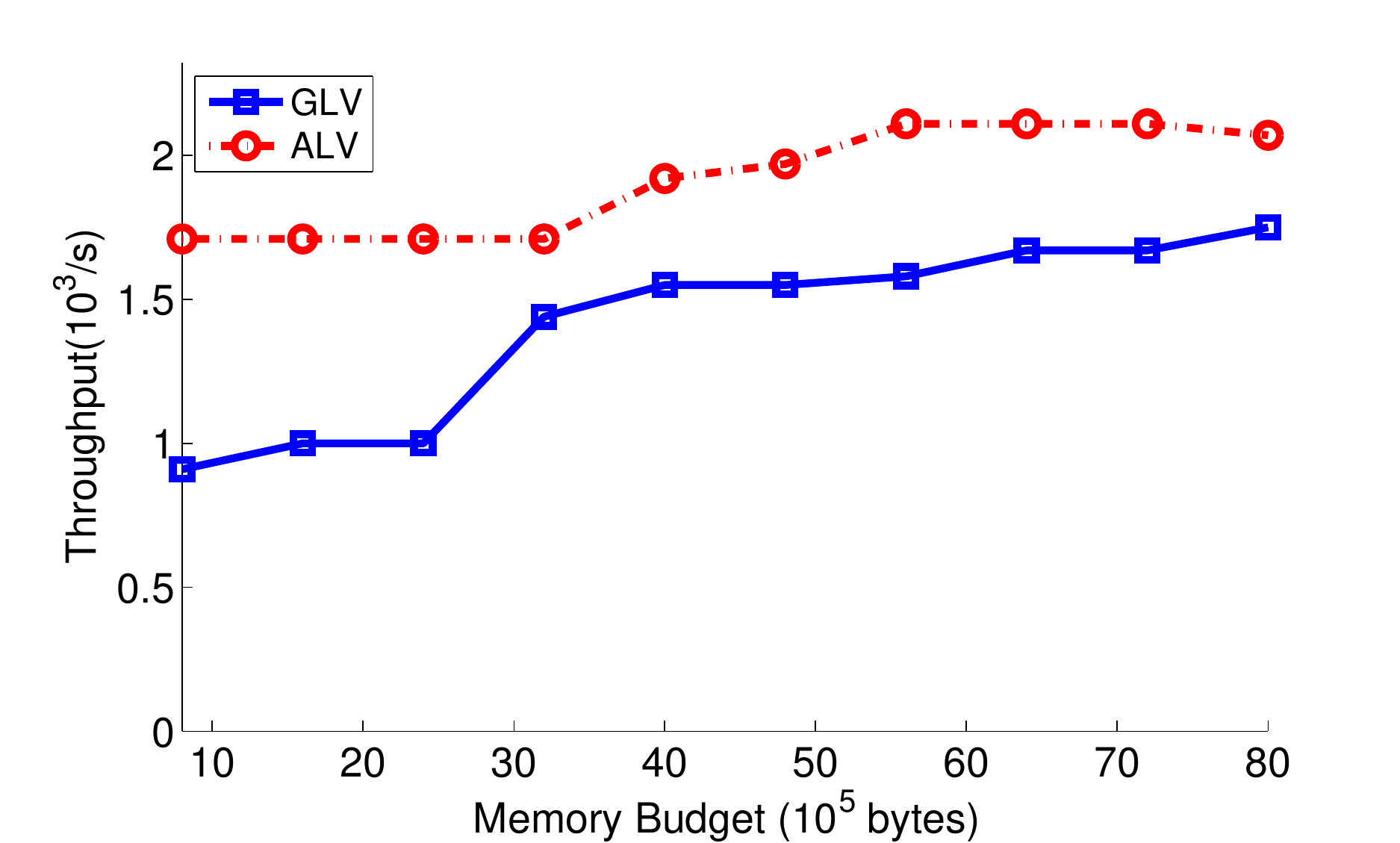}
  \label{fig:vect-1024}
}

\caption{Memory-constrained throughput of OFDM-RX systems with different 
levels of memory budget $M$ and
bandwidth $L$ using ALV-IAR compared to the GLV-HEFT baseline. 
These experiments are performed on
Intel i5-6400 CPU and NVIDIA Geforce GTX750 GPU architectures.
(a) $L=128$, (b) $L=256$, (c) $L=512$, (d) $L=1024$.}
\label{fig:ofdm-vect}
\end{figure}

The memory budgets are set to $M= b \log(L) \times 10^5$, where $b =
\{1, 2, \ldots, 10\}$.  We compare the throughput levels of implementations
generated using the two methods --- ALV-IAR and GLV-HEFT --- as shown in
Figure~\ref{fig:ofdm-vect}.  The results shown in Figure~\ref{fig:ofdm-vect}
show that using actor-level vectorization and $\mxsgm$ scheduling, we are able
to obtain system throughput that consistently exceeds 
that provided by the baseline method under
same memory constraint. 

\begin{table}
\caption{Vectorization degrees and mapping results generated
by ALV-IAR and GLV-HEFT under
the memory constraint 
$M=2.8$ Mb, and $L=512$. \label{tab:conf}}{%
\begin{tabular}{|c|c|c|c|}
\hline
Method & Vectorization & Mapping & $\mxth$($10^3/s$) \\
\hline
ALV-IAR & [1,3,12,1,1,1,1] & [0,1,1,0,0,0,0] & 3.15 \\
\hline
GLV-HEFT& [4,4,4,36,36,36,144] & [0,1,1,1,1,0,0] & 2.60 \\
\hline
\end{tabular}}
\end{table}

When memory constraints are relatively tight, GLV has difficulty in adequately
exploiting data parallelism in the OFDM-RX system.  ALV-IAR alleviates this
problem by focusing memory resources to vectorize selected,
performance-critical actors. Specifically, ALV-IAR successfully identifies
$\mxsyn$ and $\mxcfo$ as the two actors that benefit the most from vectorized
execution on the GPU. Prioritizing the vectorization of these two actors helps
to avoid wasting memory on vectorizations that have relatively little or no
impact on overall system performance. This is reflected by a large throughput
gain when $b \leq 4$. When the memory constraint is relaxed, the gap in the
throughput gain between ALV-IAR and GLV is reduced, as data-parallelism in the
system can exploited more effectively by GLV under loose memory constraints. 

When optimizing the OFDM-RX system, ALV-IAR maps only $\mxsyn$ and $\mxcfo$ onto
the GPU, and assigns the other actors to the CPU to utilize pipeline
parallelism in the system. Under this mapping, firings of $\mxsyn$ and $\mxcfo$
from subsequent frames can be executed in parallel with firings of $\mxrcp$, $\mxfft$,
$\mxdmp$ and $\mxsnk$ from earlier frames. 

In these experiments, the maximum measured speedup values of ALV-IAR over
GLV-HEFT are 2.66X, 2.45X, 1.94X and 1.71X for $L=128,256,512,1024$,
respectively. The maximum speedup values of ALV-IAR compared to a single-core,
unvectorized CPU baseline implementation are 11.1X, 19.8X, 33.8X, and 47.6X,
for $L=128,256,512,1024$, respectively. 

Although the speedup gain of ALV-IAR over GLV-HEFT in this application is
significantly higher than the average speedup in Section~\ref{sec:vect-res}, 
it still
falls within the range of the maximum speedup reported in 
Section~\ref{sec:vect-res}.
We expect that this is attributable to the relatively simple, chain-structured
topology of the application's dataflow graph. 

\begin{figure}[]
\centering

\subfigure[]{
  \includegraphics[width=2in]
  {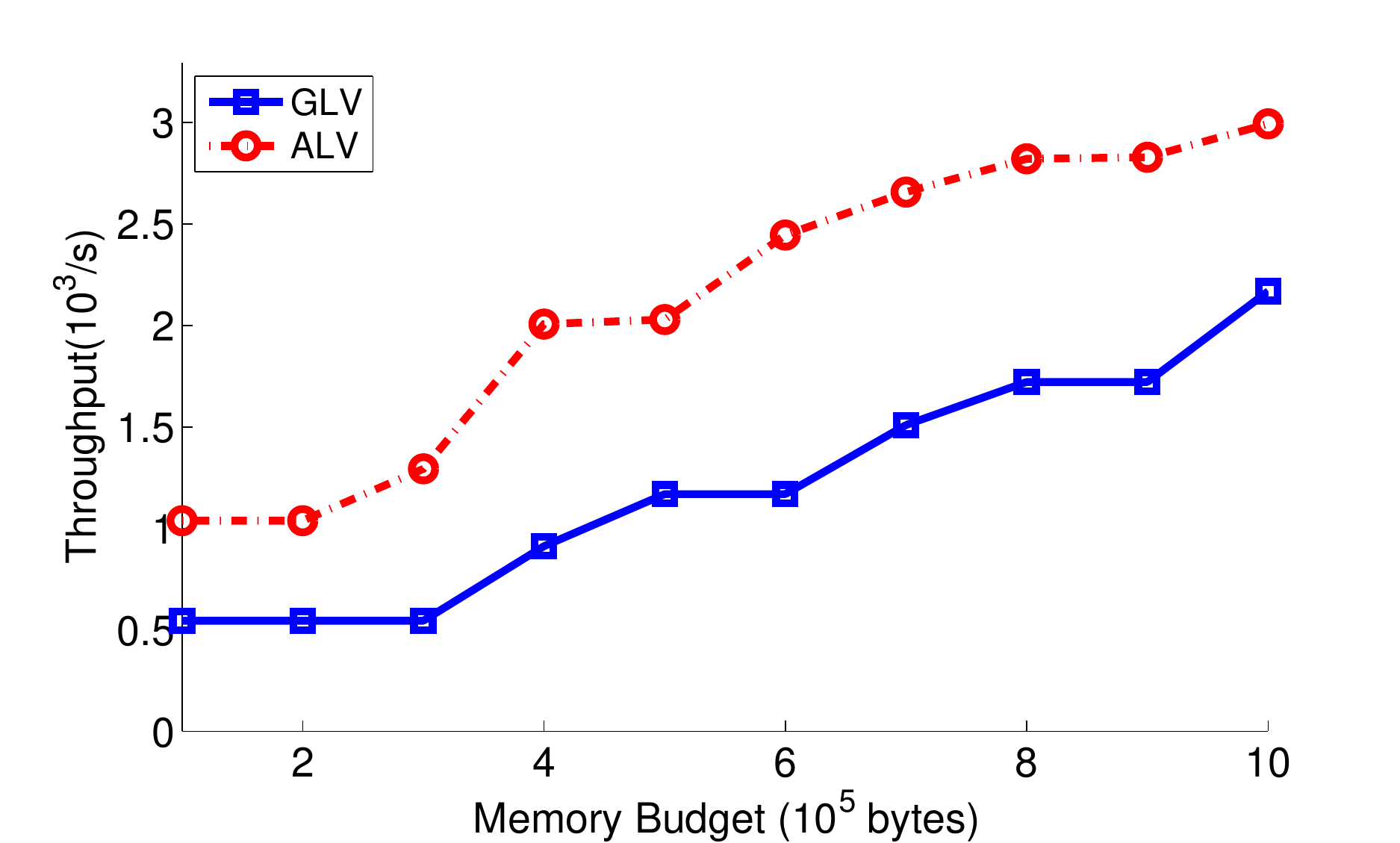}
  \label{fig:tx-vect-128}
}
\subfigure[]{
  \includegraphics[width=2in]
  {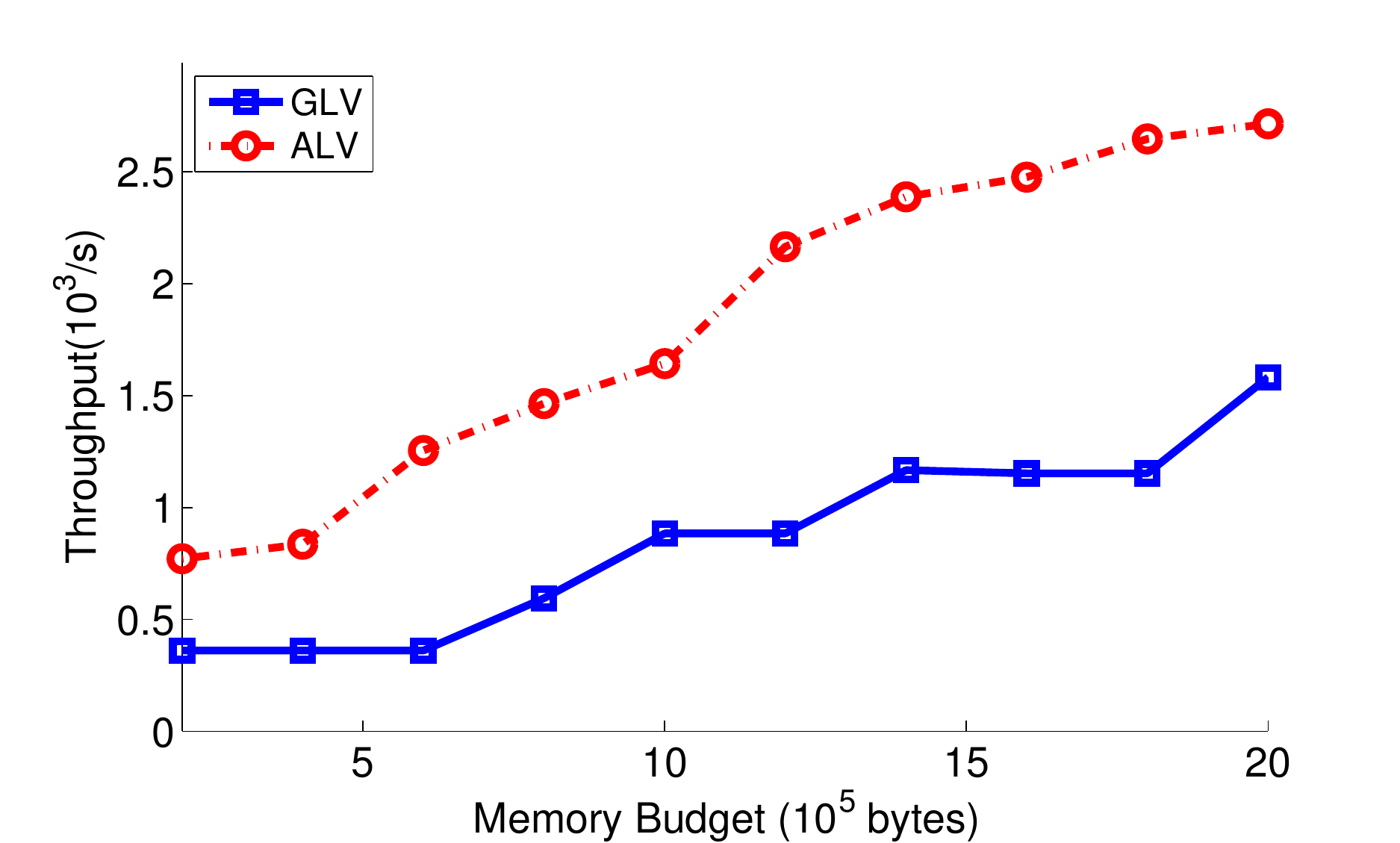}
  \label{fig:tx-vect-256}
}
\subfigure[]{
  \includegraphics[width=2in]
  {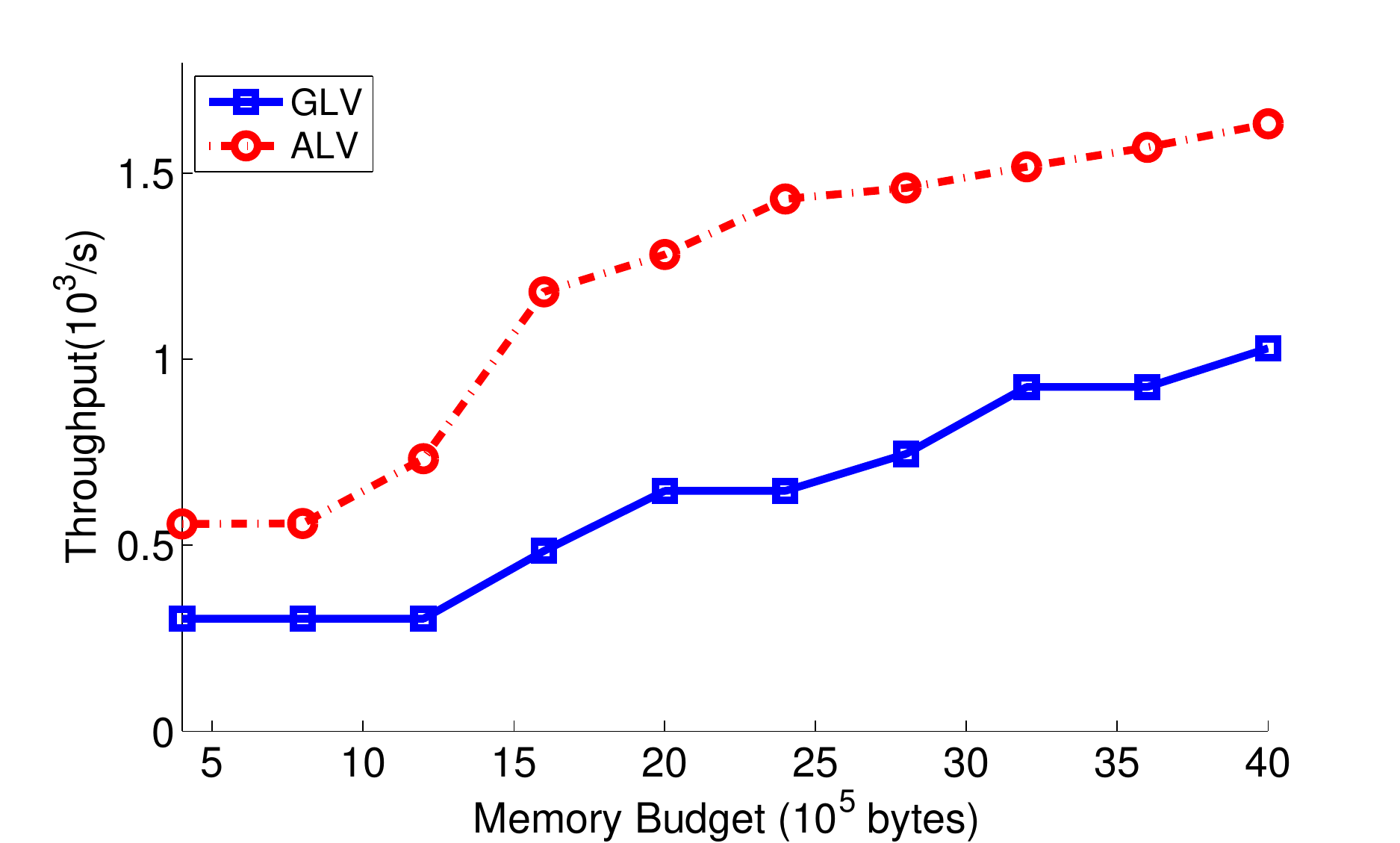}
  \label{fig:tx-vect-512}
}
\subfigure[]{
  \includegraphics[width=2in]
  {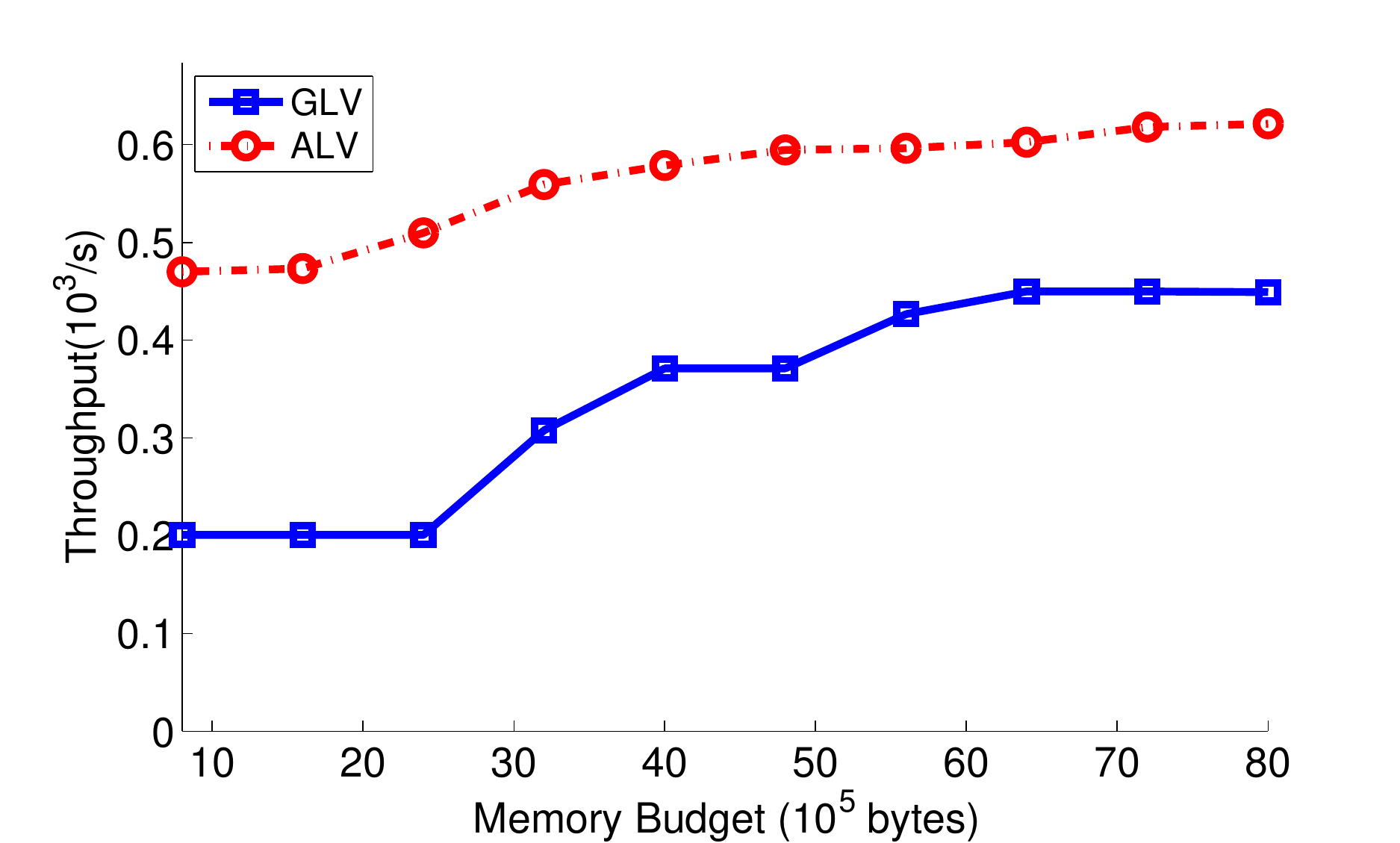}
  \label{fig:tx-vect-1024}
}

\caption{Memory-constrained throughput of OFDM-RX systems with different 
levels of memory budget $M$ and
bandwidth $L$ using ALV-IAR compared to the GLV-HEFT baseline on 
the NVIDIA Jetson TX1.  
(a) $L=128$, (b) $L=256$, (c) $L=512$, (d) $L=1024$.}
\label{fig:ofdm-tx-vect}
\end{figure}

The measurements described above are carried out on a CPU-GPU architecture in a
desktop computer platform. To complement these experiments using an embedded
platform, we investigate the performance of ALV-IAR and GLV-HEFT by performing
the same experiments on an NVIDIA Jetson TX1 (TX1). The TX1 is a popular
embedded platform that consists of a Quad-core ARM A57 CPU and an NVIDIA
Maxwell GPU with 256 CUDA cores. The results are summarized in
Figure~\ref{fig:ofdm-tx-vect}.  These results are found to be similar to those
obtained using the desktop platform.  More specifically, the maximum speedup
values of ALV-IAR over GLV-HEFT are 2.4X, 3.5X, 2.4X, 2.5X for
$L=128,256,512,1024$, respectively, as measured on the TX1. These results show
that ALV-IAR also consistently outperforms GLV-HEFT by a significant margin on
the TX1.

In summary, the throughput improvement obtained by HCGP acceleration using the
methods developed in this work facilitates real-time, memory constrained 
processing of OFDM signals. Such acceleration can benefit a variety of 
software-defined radio and cognitive radio applications. 

\section{Conclusion}
\label{sec:conclusion}

In this paper, we have investigated memory-constrained, throughput optimization
for synchronous dataflow (SDF) graphs on heterogeneous CPU-GPU platforms. We
have developed novel methods for Integrated Vectorization and Scheduling (IVS)
that provide throughput- and memory-efficient implementations on the targeted
class of platforms.  We have integrated these IVS methods into the DIF-GPU
Framework, which provides capabilities for automated synthesis of GPU software
from high-level dataflow graphs specified using the dataflow interchange format
(DIF).  Our development of novel IVS methods and their integration into DIF-GPU
provide a streamlined workflow for automated exploitation of pipeline, data
and task level parallelism from SDF graphs. We have demonstrated our IVS
methods through extensive experiments involving a large collection of diverse,
synthetic SDF graphs, as well as on a practical embedded signal processing case
study involving a wireless communications receiver that is based on orthogonal
frequency division multiplexing.  The results of our experiments demonstrate
that our proposed new methods for IVS provide significant improvements in
system throughput when mapping SDF graphs onto CPU-GPU platforms. 
Our proposed methods provide a range of useful trade-offs
between analysis time and speedup improvement that designers
can select among depending on their specific preferences and constraints.

\section{Acknowledgments}

This research was supported in part by the Laboratory for
Telecommunication Sciences and the National Science Foundation.

\bibliographystyle{ACM-Reference-Format}
\bibliography{ref}

\end{document}